\documentclass[prc,aps,twocolumn,showpacs,amssymb,superscriptaddress,fleqn]{revtex4}

\usepackage{amsmath}
\usepackage{color}
\usepackage{graphicx}
\usepackage{dcolumn}
\usepackage{mathtools}
\usepackage{relsize}
\usepackage{float}
\usepackage{braket}

\begin{document}

\title{Renormalization of the GT operator within the realistic shell model}

\author{L. Coraggio}
\affiliation{Istituto Nazionale di Fisica Nucleare, \\
Complesso Universitario di Monte  S. Angelo, Via Cintia - I-80126 Napoli, Italy}
\author{L. De Angelis}
\affiliation{Istituto Nazionale di Fisica Nucleare, \\
Complesso Universitario di Monte  S. Angelo, Via Cintia - I-80126 Napoli, Italy}
\author{T. Fukui}
\affiliation{Istituto Nazionale di Fisica Nucleare, \\
Complesso Universitario di Monte  S. Angelo, Via Cintia - I-80126 Napoli, Italy}
\author{A. Gargano}
\affiliation{Istituto Nazionale di Fisica Nucleare, \\
Complesso Universitario di Monte  S. Angelo, Via Cintia - I-80126 Napoli, Italy}
\author{N. Itaco}
\affiliation{Dipartimento di Matematica e Fisica, Universit\`a degli
  Studi della Campania ``Luigi Vanvitelli'', viale Abramo Lincoln 5 -
  I-81100 Caserta, Italy}
\affiliation{Istituto Nazionale di Fisica Nucleare, \\ 
Complesso Universitario di Monte  S. Angelo, Via Cintia - I-80126 Napoli, Italy}
\author{F. Nowacki}
\affiliation{Universit\'e de Strasbourg, IPHC, 23 rue du Loess 67037 Strasbourg, France}
\affiliation{CNRS, IPHC UMR 7178, 67037 Strasbourg, France}
\affiliation{Dipartimento di Matematica e Fisica, Universit\`a degli
  Studi della Campania ``Luigi Vanvitelli'', viale Abramo Lincoln 5 -
  I-81100 Caserta, Italy}

\begin{abstract}
In nuclear structure calculations, the choice of a limited model
space, due to computational needs, leads to the necessity to
renormalize the Hamiltonian as well as any transition operator.
Here, we present a study of the renormalization procedure and effects
of the Gamow-Teller operator within the framework of the realistic
shell model.
Our effective shell-model operators are obtained, starting from a
realistic nucleon-nucleon potential, by way of the many-body
perturbation theory in order to take into account the degrees of
freedom that are not explicitly included in the chosen model space.
The theoretical effective shell-model Hamiltonian and transition
operators are then employed in shell-model calculations, whose results
are compared with data of Gamow-Teller transition strengths and
double-$\beta$ half-lives for nuclei which are currently of interest for
the detection of the neutrinoless double-$\beta$ decay process, in a
mass interval ranging from $A=48$ up to $A=136$.
We show that effective operators are able to reproduce quantitatively
the spectroscopic and decay properties without resorting to an
empirical quenching neither of the axial coupling constant $g_A$, nor
of the spin and orbital gyromagnetic factors.
This should assess the reliability of applying present theoretical tools to
this problematic.
\end{abstract}


\maketitle

\section{Introduction}
\label{intro}
A long-standing issue of nuclear structure calculations is the need to
reduce the number of the configurations available to the interacting
nucleons, which constitute the nuclear system under investigation.
Such approximations, that are necessary to overcome the computational
complexity of the nuclear many-body problem, drive the nuclear
structure practitioners to resort to effective Hamiltonians and
operators, that depend on a certain set of parameters, built up to
account for the degrees of freedom which do not appear
explicitly in the calculated wavefunctions.
The development of effective operators suitable to describe
observables is a problematic that has to be tackled in most nuclear
structure models that relies on the truncation of the number of
interacting nucleons and/or the dimension of the configuration
space.
This issue does not affect {\it ab initio} approaches when
  their results are convergent with respect to the truncation of the
  nuclear correlations that is needed to solve the many-body
  Schr\"odinger equation (see for example
  Ref. \cite{HjorthJensen17}).

In the nuclear shell model (SM), the physics of a certain nuclear system is
described in terms only of a limited number of valence nucleons, that
interact in a model space consisting of a major shell, placed
outside a closed core made up by the remaining constituent nucleons,
the latter being frozen inside a number of filled shells.

The status of the theoretical derivation of an effective shell-model
Hamiltonian ($H_{\rm eff}$), starting from a realistic nuclear
potential, has reached nowadays a notable progress, especially within
the framework of the many-body perturbation theory
\cite{Hjorth95,Coraggio12a}.
At present, realistic shell-model Hamiltonians are largely employed in
shell-model calculations, exhibiting a substantial reliability
(see, for example Ref. \cite{Coraggio09a} and references therein).

As regards the theoretical efforts to derive effective shell-model
transition operators starting from realistic potentials, the literature is far
less extended, but it is worth mentioning an early review about
this topic, which can be found in Ref. \cite{Ellis77}.
More recently, Suzuki and Okamoto have developed a formalism to derive
effective shell-model operators \cite{Suzuki95}, that provides an
approach that is consistent with the construction of the corresponding
$H_{\rm eff}$.

In the present work we focus on the derivation of effective
shell-model Gamow-Teller (GT) operators to calculate observables
related to the $\beta$-decay transition for nuclei in different mass
regions, aiming to trace back to the roots of the quenching of the
free value of the axial coupling constant $g_A$ in nuclear structure
calculations.

It should be mentioned that similar studies have been reported in
Refs. \cite{Siiskonen01,Holt13d}.
In Ref. \cite{Siiskonen01}, the renormalization of the GT
operator, in the form of a one-body operator, has been carried out to
study the role of the weak hadronic current in the nuclear medium.
The authors of Ref. \cite{Holt13d} have instead calculated nuclear
matrix elements of the two-neutrino double-$\beta$ decay
($2\nu\beta\beta$) building an effective two-body operator within the
so-called closure approximation.

As a matter of fact, effective GT operators are in general obtained
resorting to effective values of $g_A$, via a quenching factor $q$,
to reproduce experimental GT transitions.
The choice of $q$ depends obviously on the nuclear structure model
employed to derive the nuclear wave functions, the dimensions of
the considered Hilbert space, and the mass of the nucleus under
investigation.

This problem, which affects the calculations of the Gamow-Teller
transition strengths and double-$\beta$ half-lives, has been
investigated within different nuclear-structure models, such as the
Interacting Boson Model \cite{Barea09,Barea12,Barea13}, the
Quasiparticle Random-Phase Approximation
\cite{Simkovic08,Simkovic09,Fang11,Faessler12,Simkovic18}, and the
Shell Model
\cite{Caurier08,Menendez09a,Menendez09b,Caurier12,Horoi07,Horoi13a,Horoi13b,Neacsu15,Brown15,Iwata16}.
In this regard, it is worth mentioning the recent review paper by
Suhonen \cite{Suhonen17b}, where the quenching is discussed from the
points of view of the different methods.

For the sake of clarity, we point out that the quenching of $g_A$ is
entangled with both the renormalization of many-body correlations -
due to the truncation of the basis used to construct the wave
functions - and the corrections due to the subnucleonic structure of
the nucleons \cite{Park93,Pastore09,Piarulli13,Baroni16b}, since the
free value $g_A^{free}=1.2723$ \cite{PDB18} is obtained from the data
of the neutron decay under the assumption that the nucleons are
point-like particles.

We will show in the following that the perturbative approach to the
derivation of effective spin-dependent operators allows to reproduce
quantitatively spectroscopic and decay properties without resorting to
an empirical quenching neither of the axial coupling constant $g_A$, nor
of the spin and orbital $g$-factors.

In this connection, it is worth noting that an important contribution
to understand the quenching of $g_A$ within a microscopic framework
has been given by the studies of I. S. Towner and co-workers (see the
review paper \cite{Towner87} and references therein), who have
extensively investigated the role played by both the many-body correlations
induced by the truncation of the Hilbert space and the two-body
meson-exchange currents in the renormalization of spin-dependent
electromagnetic ($M1$) and weak (GT) operators \cite{Towner83}.

Nowadays, there is a renewed interest in the problematics of the
renormalization of the GT operator, because of its connection with the
calculation of the nuclear matrix elements (NME) of the
$0\nu\beta\beta$ decay (see for example Ref. \cite{Suhonen17a}).
In fact, the half life of such a process is expressed by:

\begin{equation}
\left[ T^{0\nu}_{1/2} \right]^{-1} = G^{0\nu} \left| M^{0\nu}
\right|^2 \langle m _{\nu} \rangle^2 ~~,
\label{0nuhalflife}
\end{equation}

\noindent
where $G^{0\nu}$ is the so-called phase-space factor, $\langle m
_{\nu} \rangle$ is the effective neutrino mass, and $M^{0\nu}$ is the
nuclear matrix element, that relates the wave functions of the parent
and grand-daughter nuclei. 
As a matter of fact, $M^{0\nu}$ can be expressed as the sum of the GT, Fermi (F),
and tensor (T) matrix elements, and depends on the axial and vector
coupling constants $g_A,g_V$ :

\begin{equation}
 M^{0 \nu} = M^{0 \nu}_{GT}- \left( \frac{g_V}{g_A} \right)^2  M^{0
   \nu}_F- M^{0 \nu}_T ~~.
\label{nme00nu}
\end{equation}

On these grounds, we focus attention on the renormalization
of the GT operator that takes into account the reduced SM model space,
without considering the corrections arising from meson-exchange
currents \cite{Towner83,Baroni16a}.
Our theoretical framework is the many-body perturbation theory
\cite{Kuo81,Suzuki95,Coraggio12a,Coraggio17a}, and, starting from a realistic
nuclear potential, we derive effective shell-model GT
operators and Hamiltonians for nuclei with mass ranging from
$A=48$ to $A=136$.
We also consider in the derivation of the one-body effective operators
the so-called ``blocking effect'', to take into account the Pauli
exclusion principle in systems with more than one valence nucleon
\cite{Ellis77}.

In Section \ref{calculations} we will sketch out a few details about
the derivation of the effective SM Hamiltonians and operators from a
realistic nucleon-nucleon ($NN$) interaction.
The results of the shell-model calculations are then reported in
Section \ref{results}.
More precisely, we compare calculated and experimental $2
\nu\beta\beta$-decay matrix elements, GT transition-strength
distributions for nuclei that are candidates for the detection of the
$0 \nu\beta\beta$-decay.
We extend this analysis to magnetic dipole moments and reduced
transition probabilities ($B(M1)$), and, for the sake of completeness,
the energy spectra and $B(E2)$ values of the parent and grand-daughter
nuclei are also shown.
The conclusions of this study are drawn in Section \ref{conclusions},
together with the outlook of our current project.
In Appendix, tables containing the calculated SP energies of the
effective Hamiltonians and the matrix elements of the effective $M1$
and GT operators are reported.

\section{Outline of calculations}\label{calculations}
The cornerstone of a realistic shell-model calculation is the
choice of a realistic nuclear potential to start with.
We consider for our calculations the high-precision CD-Bonn
$NN$ potential \cite{Machleidt01b}, whose non-perturbative behavior
requires to integrate out its repulsive high-momentum components by
way of the so-called $V_{\rm low-k}$ approach
\cite{Bogner01,Bogner02}.
This is based on a unitary transformation that provides a softer
nuclear potential defined up to a cutoff $\Lambda$, and preserves the
physics of the original CD-Bonn interaction.

As in our recent works
\cite{Coraggio15a,Coraggio15b,Coraggio16a,Coraggio17a}, the value of
the cutoff $\Lambda$ is chosen equal to 2.6 fm$^{-1}$, since we have
found that the role of missing three-nucleon force (3NF) decreases by
enlarging the $V_{\rm low-k}$ cutoff \cite{Coraggio15b}.
In our experience, $\Lambda=2.6$ fm$^{-1}$ is an upper limit, since
with a larger cutoff the order-by-order behavior of the perturbative
expansion may be not satisfactory.

This $V_{\rm low-k}$ is then employed as the two-body interaction term
of the Hamiltonian for the system of $A$ nucleons under investigation:

\begin{equation}
 H =  \sum_{i=1}^{A} \frac{p_i^2}{2m} + \sum_{i<j=1}^{A} V_{\rm low-k}^{ij}
 = T + V_{\rm low-k} ~~.\label{htotal}
\end{equation}

\noindent
This Hamiltonian should be then diagonalized in an infinite Hilbert
space to describe the physical observables.
Obviously, this task is unfeasible, and in the shell model the
infinite number of degrees of freedom is reduced only to those
characterizing the physics of a limited number of interacting
nucleons, that are constrained in a finite Hilbert space spanned by a
few accessible orbitals.
To this end, the Hamiltonian $H$ of Eq. \ref{htotal} is broken up, by
way of an auxiliary one-body potential $U$, into the sum of a
one-body term $H_0$, whose eigenvectors set up the shell-model basis,
and a residual interaction $H_1$:

\begin{eqnarray}
 H &= & T + V_{\rm low-k} = (T+U)+(V_{\rm low-k}-U)= \nonumber \\
~& = &H_{0}+H_{1}~~.\label{smham}
\end{eqnarray}

The following step is to derive an effective shell-model
Hamiltonian $H_{\rm eff}$, that takes into account the degrees
of freedom that are not explicitly included in the shell-model
framework, as the core polarization due to the interaction, within
the full Hilbert space, between the valence nucleons and those belonging
to the closed core.

We derive $H_{\rm eff}$ by resorting to the many-body perturbation
theory, an approach that has been developed by Kuo and coworkers
through the 1970s \cite{Kuo90,Kuo95}.
More precisely, we use the well-known $\hat{Q}$
box-plus-folded-diagram method \cite{Kuo71}, where the $\hat{Q}$ box
is defined as a function of the unperturbed energy $\epsilon$ of the valence particles:
\begin{equation}
\hat{Q}(\epsilon) = P H_1 P + P H_1 Q \frac{1}{\epsilon - QHQ} Q H_1 P~~,
\label{qbox}
\end{equation}

\noindent
where the operator $P$ projects onto the model space and $Q=\mathbf{1}
-P$.
In the present calculations the $\hat{Q}$ box is expanded as a
collection of one- and two-body irreducible valence-linked Goldstone
diagrams up to third order in the perturbative
expansion\cite{Coraggio10a,Coraggio12a}.

Within this framework the effective Hamiltonian $H_{\rm eff}$ can be
written in an operator form as

\begin{equation}
H_{\rm eff} = \hat{Q} - \hat{Q'} \int \hat{Q} + \hat{Q'} \int \hat{Q} \int
\hat{Q} - \hat{Q'} \int \hat{Q} \int \hat{Q} \int \hat{Q} + ~...~~,
\label{heff}
\end{equation}

\noindent
where the integral sign represents a generalized folding operation
\cite{Brandow67}, and $\hat{Q'}$ is obtained from $\hat{Q}$ by
removing first-order terms \cite{Krenciglowa74}.

Since it has been demonstrated the following operatorial identity
\cite{Krenciglowa74}:

\begin{equation}
\hat{Q} \int \hat{Q}= - \hat{Q}_1\hat{Q}~~,
\end{equation}

\noindent
the solution of Eq. \ref{heff} may be obtained using the $\hat{Q}$ box derivatives

\begin{equation}
\hat{Q}_m = \frac {1}{m!} \frac {d^m \hat{Q} (\epsilon)}{d \epsilon^m} \biggl| 
_{\epsilon=\epsilon_0} ~~, 
\label{qm}
\end{equation}

\noindent
$\epsilon_0$ being the model-space eigenvalue of the unperturbed
Hamiltonian $H_0$, that we have chosen to be harmonic-oscillator
(HO) one.

Consequently, the expression in Eq. \ref{heff} may be rewritten as

\begin{equation}
H_{\rm eff} = \sum_{i=0}^{\infty} F_i~~,
\label{kkeq}
\end{equation}

\noindent
where

\begin{eqnarray}
F_0 & = & \hat{Q}(\epsilon_0)  \nonumber \\
F_1 & = & \hat{Q}_1(\epsilon_0)\hat{Q}(\epsilon_0)  \nonumber \\
F_2 & = & \left[ \hat{Q}_2(\epsilon_0)\hat{Q}(\epsilon_0) + 
\hat{Q}_1(\epsilon_0)\hat{Q}_1(\epsilon_0) \right] \hat{Q}(\epsilon_0)  \nonumber \\
~~ & ... & ~~ 
\label{kkeqexp}
\end{eqnarray}

From $H_{\rm eff}$ for one-valence-nucleon systems we obtain the
single-particle (SP) energies for our SM calculations, while the
two-body matrix elements (TBMEs) are obtained from $H_{\rm eff}$
derived for the nuclei with two valence nucleons, by subtracting the
theoretical SP energies.
The calculated SP energies for $^{40}$Ca, $^{56}$Ni, and $^{100}$Sn
cores are reported in the Appendix, while the corresponding TBMEs can
be found in the Supplemental Material \cite{supplemental2018}.

A detailed description of the perturbative properties of our $H_{\rm
  eff}$, derived from the same  $V_{\rm low-k}$ of present work, can
be found in \cite{Coraggio18b}, where it has been reported the
behavior of SP energies and TBME as a function of both the
perturbative order and the number of the intermediate states.

\begin{center}
\begin{figure}[ht]
\includegraphics[scale=0.36,angle=0]{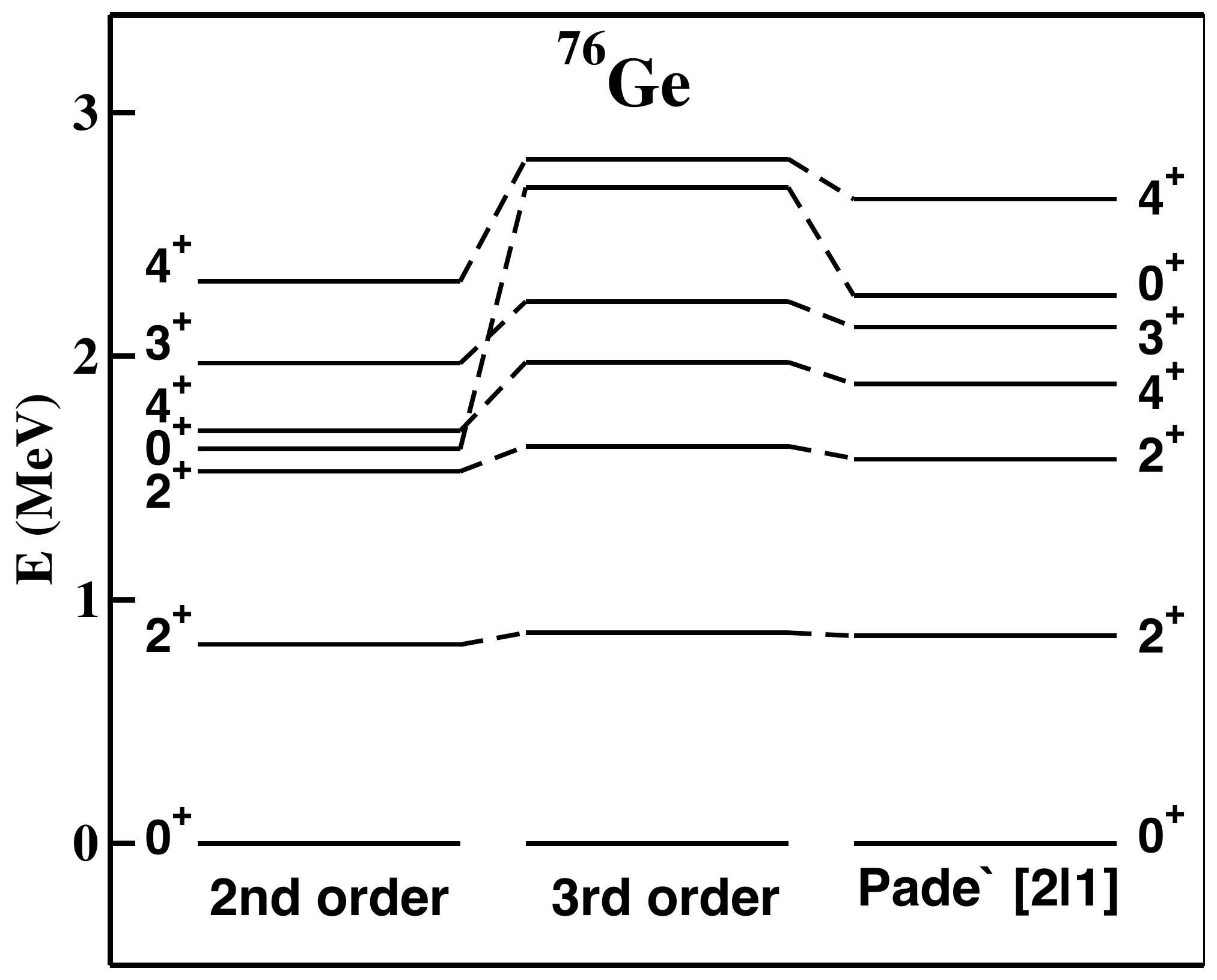}
\caption{Calculated spectra of $^{76}$Ge obtained from the
  perturbative expansion of $H_{\rm eff}$ using $\hat{Q}$-box diagrams
  up to second-, third-order, and Pad\'e approximant $[2|1]$ (see text
  for details).}
\label{76Ge_conv}
\end{figure}
\end{center}

  In order to exemplify pictorially the impact of the perturbation
  expansion on the energy spectra, we report in Fig. \ref{76Ge_conv}
  the low-energy spectra of $^{76}$Ge obtained with $H_{\rm eff}$s
  derived from $\hat{Q}$-boxes at second-, third-order in perturbation
  theory, and their Pad\'e approximant $[2|1]$ \cite{Baker70}.
  We employ the Pad\'e approximant in order to obtain a better
  estimate of the convergence value of the perturbation series
  \cite{Coraggio12a}, as suggested in \cite{Hoffmann76}.
  As can be seen, a rather good convergence is obtained, apart from
  the second-excited $J^{\pi}=0^+$ state, the largest discrepancy
  occurring for the yrare $J^{\pi}=4^+$ that is about $20\%$ from
  second to third order and $6\%$  from third order to Pad\'e
  approximant $[2|1]$.

As mentioned before, we derive the effective transition operators,
namely the matrix elements of the effective spin-dependent $M1$, GT
operators and the effective charges of the electric quadrupole
operator, using the formalism presented by Suzuki and Okamoto in
Ref. \cite{Suzuki95}.

As a matter of fact, a non-Hermitian effective operator $\Theta_{\rm
  eff}$ can be expressed in terms of the $\hat{Q}$ box, its
derivatives, and an infinite sum of operators $\chi_n$, the latter
being defined as:

\begin{eqnarray}
\chi_0 &=& (\hat{\Theta}_0 + h.c.)+ \hat{\Theta}_{00}~~,  \label{chi0} \\
\chi_1 &=& (\hat{\Theta}_1\hat{Q} + h.c.) + (\hat{\Theta}_{01}\hat{Q}
+ h.c.) ~~, \\
\chi_2 &=& (\hat{\Theta}_1\hat{Q}_1 \hat{Q}+ h.c.) +
(\hat{\Theta}_{2}\hat{Q}\hat{Q} + h.c.) + \nonumber \\
~ & ~ & (\hat{\Theta}_{02}\hat{Q}\hat{Q} + h.c.)+  \hat{Q}
\hat{\Theta}_{11} \hat{Q}~~, \label{chin} \\
&~~~& \cdots \nonumber
\end{eqnarray}

\noindent
where $\hat{\Theta}_m$, $\hat{\Theta}_{mn}$ have the following
expressions:
\begin{eqnarray}
\hat{\Theta}_m & = & \frac {1}{m!} \frac {d^m \hat{\Theta}
 (\epsilon)}{d \epsilon^m} \biggl|_{\epsilon=\epsilon_0} ~~~, \\
\hat{\Theta}_{mn} & = & \frac {1}{m! n!} \frac{d^m}{d \epsilon_1^m}
\frac{d^n}{d \epsilon_2^n}  \hat{\Theta} (\epsilon_1 ;\epsilon_2)
\biggl|_{\epsilon_1= \epsilon_0, \epsilon_2  = \epsilon_0} ~,
\end{eqnarray}

\noindent
with
\begin{eqnarray}
\hat{\Theta} (\epsilon) = & P \Theta P + P \Theta Q
\frac{1}{\epsilon - Q H Q} Q H_1 P ~, ~~~~~~~~~~~~~~~~~~~\label{thetabox} \\
\hat{\Theta} (\epsilon_1 ; \epsilon_2) = & P H_1 Q
\frac{1}{\epsilon_1 - Q H Q} Q \Theta Q \frac{1}{\epsilon_2 - Q H Q} Q H_1 P ~,~~~~~
\end{eqnarray}

\noindent
$\Theta$ being the bare transition operator.

The effective transition operators can be written, in terms of the
above quantities, as follows

\begin{eqnarray}
\Theta_{\rm eff} & = & (P + \hat{Q}_1 + \hat{Q}_1 \hat{Q}_1 + \hat{Q}_2
\hat{Q} + \hat{Q} \hat{Q}_2 + \cdots)\nonumber \\
~ & ~& \times (\chi_0+ \chi_1 + \chi_2 +\cdots)~~. \label{effopexp1}
\end{eqnarray}

Now, inserting the identity $\hat{Q}  \hat{Q}^{-1} = \mathbf{1}$ and
taking into account Eqs. \ref{kkeq}, \ref{kkeqexp},
Eq. \ref{effopexp1} may be then recast in the following form

\begin{eqnarray}
\Theta_{\rm eff} & = & (P + \hat{Q}_1 + \hat{Q}_1 \hat{Q}_1 + \hat{Q}_2
\hat{Q} + \hat{Q} \hat{Q}_2 + \cdots) \hat{Q}  \hat{Q}^{-1} \nonumber \\
~ & ~& \times (\chi_0+ \chi_1 + \chi_2 +\cdots) = \nonumber \\
~ & = & H_{\rm eff} \hat{Q}^{-1}  (\chi_0+ \chi_1 + \chi_2 +\cdots) ~~,
\label{effopexp2}
\end{eqnarray}

The above form provides a strong link between the derivation of the
effective Hamiltonian and all effective operators.

In our calculations for the aforementioned one-body transition
operators, we arrest the $\chi_n$ series to the $\chi_2$ term. 
It is worth reminding that in Refs. \cite{Coraggio17a, Coraggio18b} we
have included only the leading term $\chi_0$.
The calculation is performed starting from a perturbative expansion of
$\hat{\Theta}_0$ and $\hat{\Theta}_{00}$, including diagrams up to the
third order in the perturbation theory, consistently with the
perturbative expansion of the $\hat{Q}$ box.
We have found that $\chi_2$ contribution is at most $1\%$ of the final
results.
Since $\chi_3$ depends on the first, second, and third derivatives of
$\hat{\Theta}_0$ and $\hat{\Theta}_{00}$, and on the first and second
derivatives of the $\hat{Q}$ box (see Eq. \ref{chin}), our estimation
of these quantities leads to evaluate $\chi_3$ being at least one
order of magnitude smaller than $\chi_2$.

\begin{figure}[ht]
\begin{center}
\includegraphics[scale=0.40,angle=0]{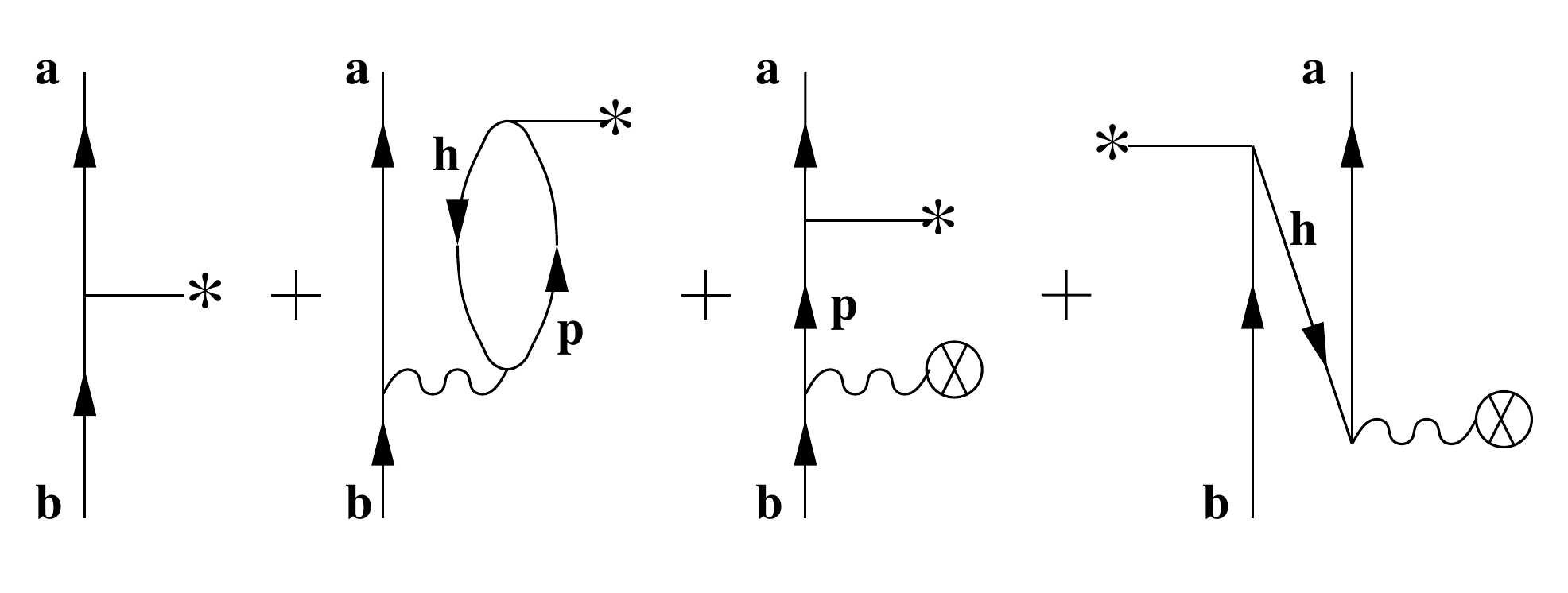}
\caption{One-body second-order diagrams included in the perturbative
  expansion of $\hat{\Theta}$. The asterisk indicates the bare operator
  $\Theta$, the wavy lines the two-body potential $V_{\rm low-k}$.}
\label{figeffop1}
\end{center}
\end{figure}

For the sake of clarity, in Fig. \ref{figeffop1} we report all the
one-body $\Theta_0$ diagrams up to the second order, the bare operator
$\Theta$ being represented with an asterisk.
The first-order $(V_{\rm low-k}-U)$-insertion, represented by a
circle with a cross inside, arises because of the presence of the $−U$
term in the interaction Hamiltonian $H_1$ (see for example
Ref. \cite{Coraggio12a} for details).

In Ref. \cite{Coraggio18b} we have carried out a study of the
perturbative properties of the GT operator for the calculation of the
NME ($M_{\rm GT}^{2\nu}$) of the $^{130}$Te, $^{136}$Xe
$2\nu\beta\beta$ decay.
We have found that the results for the $^{130}$Te decay vary by about
$10\%$ from second to third order in perturbation theory, and that for
$^{136}$Xe by about $5\%$.

As regards the magnetic-dipole operator, we find a similar
perturbative behavior.
As a matter of fact, the calculated magnetic dipole moments of the
yrast $J^\pi = 2^+$ states in $^{130}$Te and $^{136}$Xe, obtained with
an effective operator derived at second order in perturbation theory,
are 0.65 and 1.19 $\mu_N$, respectively, to be compared with 0.71 and
1.15 $\mu _N$ at third-order (see also Tables
\ref{M1_A130},\ref{M1_A136}.)
The variation from second to third order is about $8\%$ for
$^{130}$Te, and $3\%$ for $^{136}$Xe.

The authors of Ref. \cite{Siiskonen01} have carried out
  a perturbative expansion of the $\hat{\Theta}$ operator for GT
  transitions in terms of similar diagrams, calculated up to
  third-order in perturbation theory and employing $G$-matrix
  energy-dependent interaction vertices.
They have added to them the corresponding folded diagrams according to
the prescription of Ref. \cite{Towner87}, but have  neglected all
diagrams with $(G-U)$-insertion vertices, which -  it is worth to point
out - at second order only are equal to zero for spin-dependent
operators.

They have reported a selection of the matrix elements of their
effective GT$^+$ operator, that we compare with our results in Tables
\ref{effGT_40Ca},\ref{effGT_56Ni}, and \ref{effGT_100Sn}.
As can be observed, both calculations provide consistent results, even
if some matrix elements differ up to $25\%$.

\begin{figure}[ht]
\begin{center}
\includegraphics[scale=0.40,angle=0]{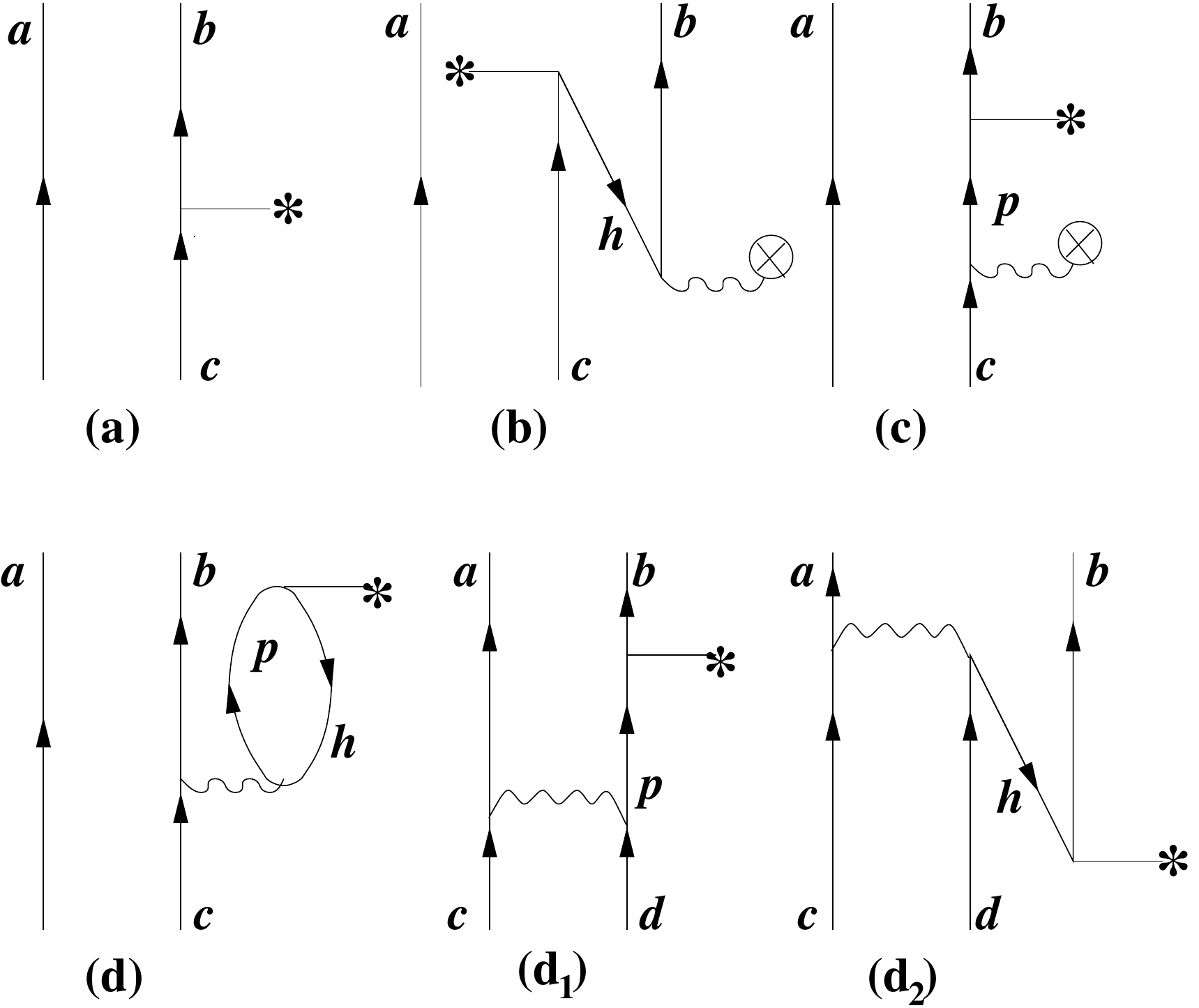}
\caption{Two-body second-order diagrams that should appear in the
  perturbative expansion of $\chi_0$. As in Fig. \ref{figeffop1}, the
  asterisk indicates the bare operator $\Theta$, the wavy lines the
  two-body potential $V_{\rm low-k}$.}
\label{figeffop2}
\end{center}
\end{figure}

The topology of the diagrams reported in Fig. \ref{figeffop1} deals,
obviously, with single-valence nucleon systems, and many-body diagrams
should be included starting from nuclei with two valence nucleons on; in
Fig. \ref{figeffop2} we report all two-valence-nucleon
diagrams for one-body operators, up to second order of the
$\hat{\Theta}$ perturbative expansion.
For the sake of simplicity, for each topology we draw only one of the
diagrams which correspond to the exchange of the external pairs of
lines.

Diagrams (a)-(d) are the same as in Fig. \ref{figeffop1} but with a
spectator line $a$, while connected diagrams ($\mathrm{d}_1$) and
($\mathrm{d}_2$) correct Pauli-principle violation introduced by
diagram (d) when the particles $c$ and $p$ own the same quantum
numbers.

Since it is straightforward to perform shell-model calculations using
one-body transition operators, we derive a density-dependent
one-body operator from the two-body ones by summing and averaging over
one incoming and outcoming particles of the connected diagrams
($\mathrm{d}_1$)  and ($\mathrm{d}_2$) of Fig. \ref{figeffop2}.
This allows to take into account the filling of the model-space
orbitals when dealing with more than one valence nucleon.

\begin{figure}[ht]
\begin{center}
\includegraphics[scale=0.40,angle=0]{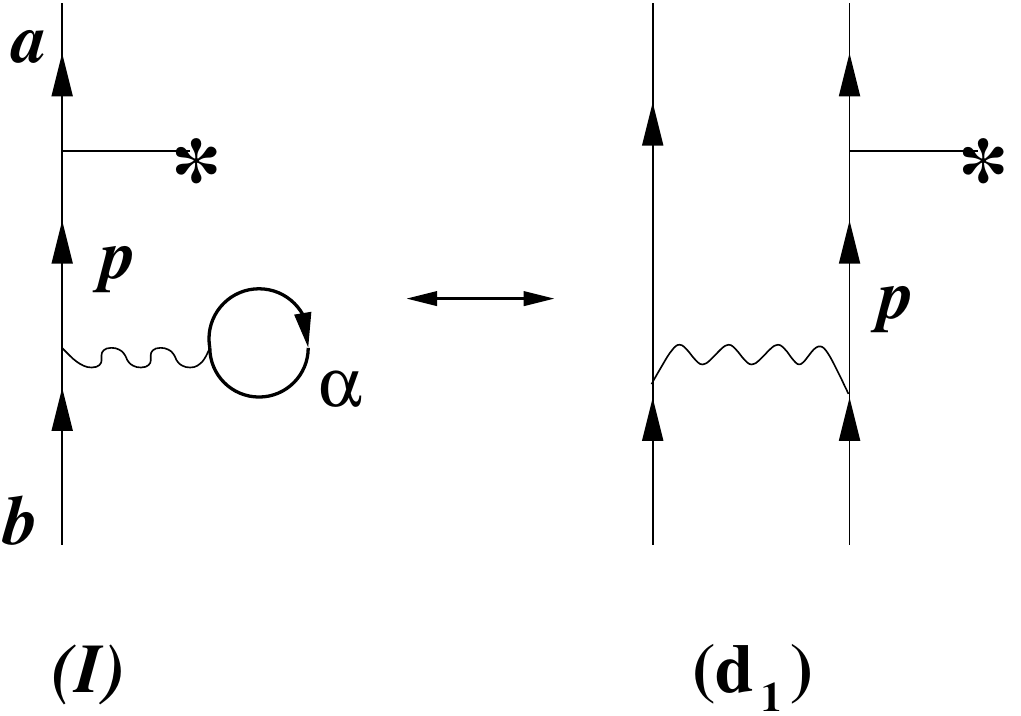}
\caption{Density-dependent one-body second-order diagram $(I)$ that is
  obtained from diagram $(\mathrm{d}_1)$ of Fig. \ref{figeffop1} by
  summing over one incoming and outcoming particles of the two-body
  diagram (see text for details).}
\label{figeffop3}
\end{center}
\end{figure}

For example, we report in Fig. \ref{figeffop3} the second-order
density-dependent one-body diagram ($I$), obtained from the
contribution ($\mathrm{d}_1$) of Fig. \ref{figeffop2}, and whose
explicit expression is

\begin{equation}
  (I) = \sum_{\alpha} \sum_{pJ} \frac{(2J+1)}{(2j_b+1)}
  \frac{\langle a || \Theta || p \rangle \langle p \alpha, J |V_{\rm low-k} | b
  \alpha , J \rangle}{ (\epsilon_b -\epsilon_p)}\rho(\alpha)~~,
\end{equation}

\noindent
where $\alpha$ and $p$ indices run over the orbitals in, and above the
model space, respectively, the matrix elements of the $V_{\mathrm
  low-k}$ are coupled to the total angular momentum $J$, $\epsilon_i$
stands for the unperturbed energy of the orbital $i$, and
$\rho(\alpha)$ is the occupation probability of the orbital $\alpha$.

In this work all the results of the shell-model calculations, that are
shown in Sec. \ref{results}, have been obtained employing SP
energies, TBMEs, and effective one-body operators derived by way of the
above mentioned theoretical approach, including consistently all
contributions up to third-order in the perturbative expansion, without
resorting to any empirically fitted parameter.

In Sec. \ref{results}, the calculated running sums of the GT strengths
($\Sigma B(p,n)$), obtained with both bare and effective GT
operators, are reported as a function of the excitation energy, and
compared with the available data extracted from experiment.
The GT strength is defined as follows:

\begin{equation}
B(p,n) = \frac{ \left| \langle \Phi_f || \sum_{j}
  \vec{\sigma}_j \tau^-_j || \Phi_i \rangle \right|^2} {2J_i+1}~~,
\label{GTstrength}
\end{equation}

\noindent
where indices $i,f$ refer to the parent and grand-daughter nuclei,
respectively, and the sum is over all interacting nucleons.

The single-$\beta$ decay GT strengths, defined by
Eq. (\ref{GTstrength}), can be accessed experimentally through
intermediate energy charge-exchange reactions, since the $\beta$-decay
process is forbidden for the nuclei under our investigation.
The GT strength can be extracted from the GT component of the cross
section at zero degree, following the standard approach in the
distorted-wave Born approximation (DWBA) \cite{Goodman80,Taddeucci87}:

\begin{equation}
\frac{d\sigma^{GT}(0^\circ)}{d\Omega} = \left (\frac{\mu}{\pi \hbar^2} \right
)^2 \frac{k_f}{k_i} N^{\sigma \tau}_{D}| J_{\sigma \tau} |^2 B(p,n)~~,
\end{equation}

\noindent
where $N^{\sigma \tau}_{D}$ is the distortion factor, $| J_{\sigma
  \tau} |$ is the volume integral of the effective $NN$ interaction,
$k_i$ and $k_f$ are the initial and final momenta, respectively, and
$\mu$ is the reduced mass.

As regards the calculation of the NME of the $2\nu\beta\beta$ decay,
it can be obtained via the following expression:

\begin{equation}
M_{\rm GT}^{2\nu}= \sum_n \frac{ \langle 0^+_f || \vec{\sigma} \tau^-
  || 1^+_n \rangle \langle 1^+_n || \vec{\sigma}
\tau^- || 0^+_i \rangle } {E_n + E_0} ~~,
\label{doublebetame}
\end{equation}

\noindent
where $E_n$ is the excitation energy of the $J^{\pi}=1^+_n$
intermediate state, $E_0=\frac{1}{2}Q_{\beta\beta}(0^+) +\Delta M$,
$Q_{\beta\beta}(0^+)$ and $\Delta M$ being the $Q$ value of the $\beta
\beta$ decay and the mass difference between the daughter and parent
nuclei, respectively.
In the above equation the index $n$ runs over all possible
intermediate states of the daughter nucleus.
The NMEs have been calculated using the ANTOINE shell-model code,
using the Lanczos strength-function method as in
Ref. \cite{Caurier05}, and including as many as intermediate states to
obtain at least a four-digit accuracy (see also Figs. 5,11 in
Ref. \cite{Coraggio17a}).
The theoretical values are then compared with the experimental
counterparts, that are extracted from the observed
half life $T^{2\nu}_{1/2}$ 

\begin{equation}
\left[ T^{2\nu}_{1/2} \right]^{-1} = G^{2\nu} \left| M_{\rm GT}^{2\nu}
\right|^2 ~~.
\label{2nihalflife}
\end{equation}

The calculation of $M_{\rm GT}^{2\nu}$ may be also performed without
calculating explicitly the intermediate $J^{\pi}=1^+_n$ states of the
daughter nucleus, namely resorting to the so-called closure
approximation \cite{Haxton84}.
The price to be payed is that the transition operator is no longer a
one-body operator but a two-body one.
It is worth pointing out that this approximation is largely employed
to calculate the $0\nu\beta\beta$ NME ($M^{0\nu}$), since the high
momentum of the neutrino - which appears explicitly in the definition
of $M^{0\nu}$ - is about 100 MeV that is one order of magnitude
greater than the average $J^{\pi}=1^+_n$ excitation energy.
As a matter of fact, it has been estimated that this approximation is
valid within $10\%$ of the exact result \cite{Senkov13}.
Actually, the same approximation has turned out to be very
unsatisfactory for the calculation of $M_{\rm GT}^{2\nu}$, because the
energies of the neutrinos which are emitted in the $2\nu\beta\beta$
process are much smaller.
For instance, in Ref. \cite{Holt13d} the authors obtain a result for
$^{76}$Ge $M_{\rm GT}^{2\nu}$ that is about two times larger than the one
calculated with the Lanczos strength-function method by employing the
same SM wave functions \cite{Horoi13c,Brown15}, and about 5 times
larger than the experimental value \cite{Barabash15}.

\section{Results}\label{results}

In this section we present the results of our SM calculations.

We compare the calculated low-energy spectra of $^{48}$Ca,
$^{48}$Ti, $^{76}$Ge, $^{76}$Se, $^{82}$Se, $^{82}$Kr, $^{130}$Te,
$^{130}$Xe, $^{136}$Xe, and $^{136}$Ba, and their electromagnetic
properties with the available experimental counterparts.
As mentioned in the Introduction, special attention will be focussed
on the magnetic dipole properties, since both $M1$ and GT operators
are spin dependent.

We show also the results of the GT$^-$ strength distributions and the
calculated NMEs of the $2\nu\beta\beta$ decays for $^{48}$Ca,
$^{76}$Ge, $^{82}$Se, $^{130}$Te, and $^{136}$Xe, and compare them
with the available data.
All the calculations have been performed employing theoretical SP
energies, TBMEs, and effective transition operators.
In particular, for the $M1$ and GT properties we report the calculated
values obtained by using the bare (I) and effective (II) operators, as
well as those including the blocking effect - and labelled as (III) -
by way of a density-dependent effective operator as mentioned in
Sec. \ref{calculations}. 
The latter give us the opportunity to investigate the role of
many-body correlations on the spin- and spin-isospin-dependent
one-body operators in nuclei with more than one valence nucleon.
 
\subsection{$^{48}$Ca}\label{48Ca}
\begin{center}
\begin{figure}[ht]
\includegraphics[scale=0.30,angle=0]{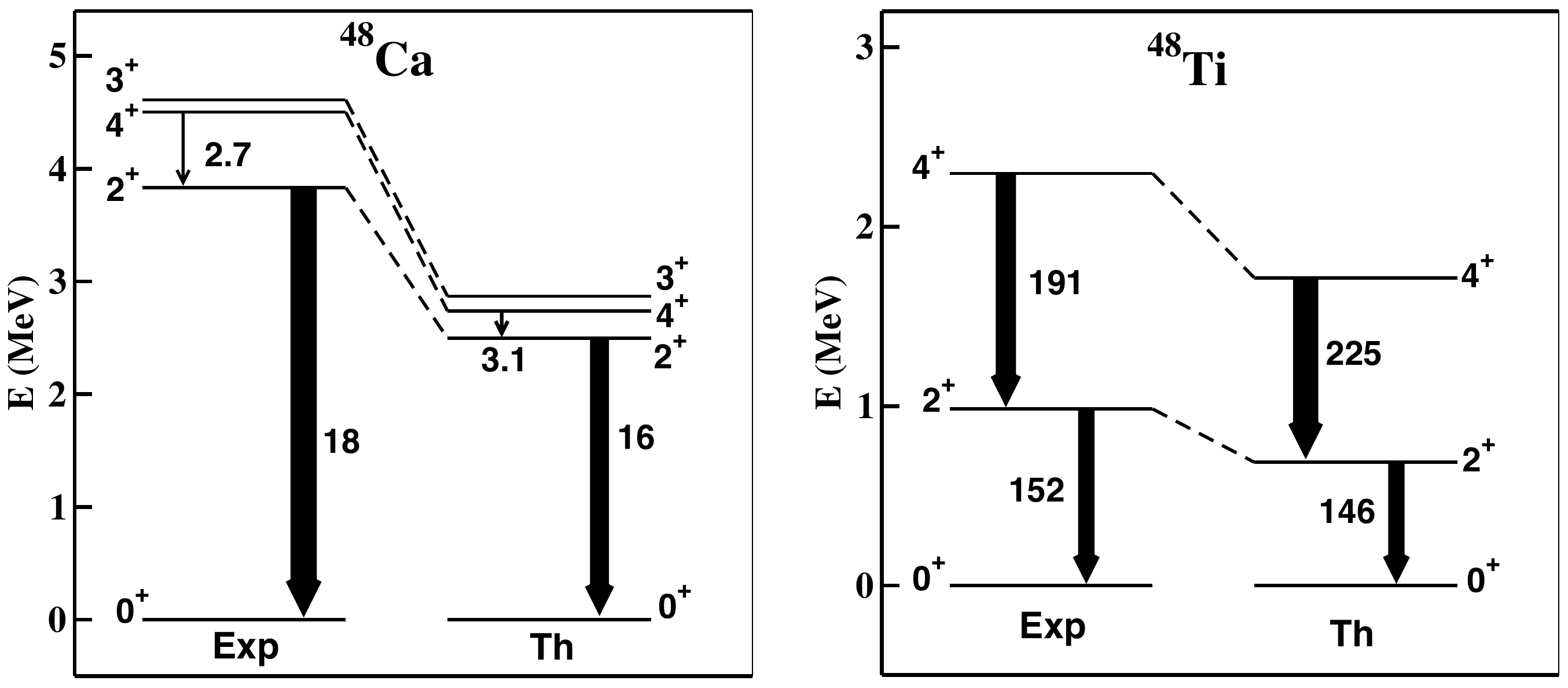}
\caption{Experimental and calculated spectra of $^{48}$Ca and
  $^{48}$Ti. $B(E2)$ strengths (in $e^2{\rm fm}^4$) are also reported
  (see text for details).}
\label{48Ca48Ti}
\end{figure}
\end{center}

The shell-model calculation for $^{48}$Ca and $^{48}$Ti are performed
within the full $fp$ shell, namely the proton and neutron $0f_{7/2}$,
$0f_{5/2}$, $1p_{3/2}$, and $1p_{1/2}$ orbitals. 
In Fig. \ref{48Ca48Ti}, we show the experimental \cite{ensdf,xundl}
and calculated low-energy spectra of $^{48}$Ca and $^{48}$Ti.
Next to the arrows, that are proportional to the $B(E2)$ strengths, we
report the explicit experimental \cite{ensdf,xundl} and calculated
$B(E2)$s in $e^2{\rm fm}^4$.

As can be seen, we do not reproduce the observed shell-closure of  the
neutron $0f_{7/2}$ orbital in $^{48}$Ca and the agreement between the
experimental and calculated spectra is only qualitative, while
experimental $B(E2)$s are satisfactorily reproduced by the theory.

In Table \ref{M1_A48} the low-energy experimental and calculated
observable related to the $M1$ operator are reported.
The calculated values reported in columns I, II, III are obtained with
the bare magnetic-dipole operator, the effective one without blocking
effect, and the one including the blocking effect, respectively.

\begin{table}[ht]
\caption{Experimental and calculated $B(M1)$ strengths (in $\mu_N^2$)
  and magnetic dipole moments (in $\mu_N$) of $^{48}$Ca and
  $^{48}$Ti. We report those for the observed states in
  Fig. \ref{48Ca48Ti} (see text for details).}
\begin{ruledtabular}
\begin{tabular}{cccccc}
\label{M1_A48}
 Nucleus & $J_i \rightarrow J_f $ & $B(M1) _{\rm Expt}$ & I & II & III \\
\colrule
 ~ & ~ & ~ & ~ & ~ & ~\\
 $^{48}$Ca   &          ~                     & ~ & ~ & ~ & ~\\
~  & $3^+_1 \rightarrow 2^+_1$ & $0.023 \pm 0.004$ \cite{Vanhoy92} & 0.051 & 0.046 & 0.046  \\
 ~ & ~ & ~ & ~ & ~ & ~\\
\colrule
 Nucleus & $J$ & $\mu _{\rm Expt}$ & I & II & III \\
\colrule
 ~ & ~ & ~ & ~ & ~ & ~\\
 $^{48}$Ti  &     ~        &           ~              &  ~   &  ~   & ~\\
        ~      & $2^+_1$ & $+0.78 \pm 0.04$ \cite{ensdf} & +0.37 & +0.54 & +0.54  \\
~              &      ~      & $+0.9 \pm 0.4$ \cite{ensdf}&    ~  &   ~   &~  \\
~              & $4^+_1$ & $+2.2 \pm 0.5$  \cite{ensdf}  & +1.2 & +1.5 & +1.5  \\
\end{tabular}
\end{ruledtabular}
\end{table}

From inspection of Table \ref{M1_A48}, it can be seen that the
calculated magnetic-dipole transition rates $B(M1)$s compare well with
the observed value for $^{48}$Ca.
In particular, from Table \ref{effM1_40Ca}, the values obtained
employing the effective shell-model operators (II-III) are quenched
with respect to that calculated with the bare operator (I), and in a
better agreement with experiment.
The blocking effect is very tiny because the number of valence
nucleons is rather small compared with the full capacity of the $fp$
shell.

As regards the magnetic moments, data are available for $^{48}$Ti, and
they are underestimated by the theory.
However, the contribution due to the effective operators points in the
right direction, leading to a better agreement with experiment.

\begin{table}[ht]
  \caption{Experimental \cite{Barabash15}  and calculated NME of the
    $2\nu\beta\beta$ decay (in MeV$^{-1}$) for $^{48}$Ca. The same
    notation of Table \ref{M1_A48} is used (see text for details).}
\begin{ruledtabular}
\begin{tabular}{ccccc}
\label{ME_48Ca}
 Decay & NME$_{\rm Expt}$ & I & II & III \\
\colrule
~ & ~ & ~ & ~ & ~\\
$^{48}$Ca  $\rightarrow$ $^{48}$Ti & $0.038 \pm 0.003$ & 0.030 & 0.026 &  0.026 \\
\colrule
\end{tabular}
\end{ruledtabular}
\end{table}

In Table \ref{ME_48Ca} we report the observed and calculated values of
the NMEs for the $2\nu\beta\beta$ decay of $^{48}$Ca into $^{48}$Ti.
The NME obtained with the bare operator (I) slightly underestimates
the experimental one, and it is $20\%$ larger than those obtained with the
effective operators (II) and (III).
This corresponds to a quenching factor $q=0.9$, that is roughly the 
average value of the reduction factor that can be extracted from Table
\ref{effGT_40Ca}, comparing the single-particle elements of the bare
GT operator with the effective ones.
In this context, it should be mentioned that our $2\nu\beta\beta$ NME
calculated with the bare operator is very different from those
obtained by way of SM calculations employing phenomenological $H_{\rm
  eff}$ \cite{Horoi07,Caurier12}, which reproduce correctly the
observed shell-closure of $0f_{7/2}$ orbital in $^{48}$Ca.

It is worth noting that, as for the M1 properties, the blocking effect
plays a negligible role also in the calculation of the $2\nu \beta
\beta$ NME.

\begin{center}
\begin{figure}[ht]
\includegraphics[scale=0.36,angle=0]{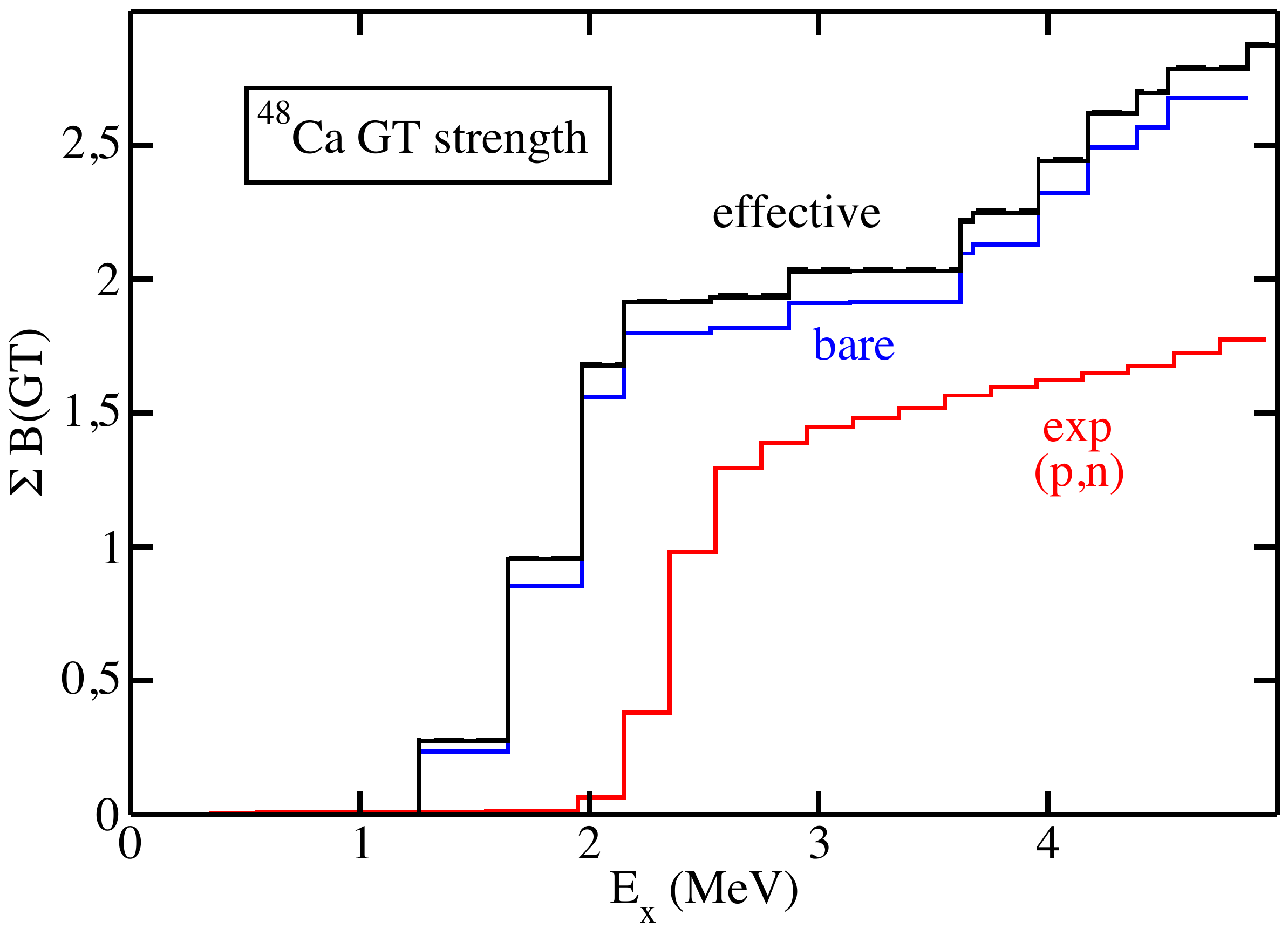}
\caption{Running sums of the $^{48}$Ca $B(p,n)$
  strengths as a function of the excitation energy $E_x$ up to 5 MeV
  (see text for details).}
\label{48CaGT-}
\end{figure}
\end{center}

In Fig. \ref{48CaGT-}, the calculated $\Sigma B(p,n)$ for
$^{48}$Ca are shown as a function of the excitation energy, and
compared with the data reported with a red line \cite{Yako09}.
The results obtained with the bare operator (I) are drawn with a blue
line, while those obtained employing the effective GT operators
without and with the blocking effect are plotted using continuous and
dashed black lines, respectively.

It can be seen that the distribution obtained using the bare operator
(I) overestimates the observed one, and it is very close to those
provided by both the effective GT operators (II-III), the blocking
effect being almost negligible.
Finally, we report about the theoretical total ${\rm GT}^-$ strengths
that are 24.0, 23.1, and 23.0 with the bare operator (I), and the
effective ones (II) and (III), respectively.

\subsection{$^{76}$Ge}\label{76Ge}
\begin{center}
\begin{figure}[ht]
\includegraphics[scale=0.30,angle=0]{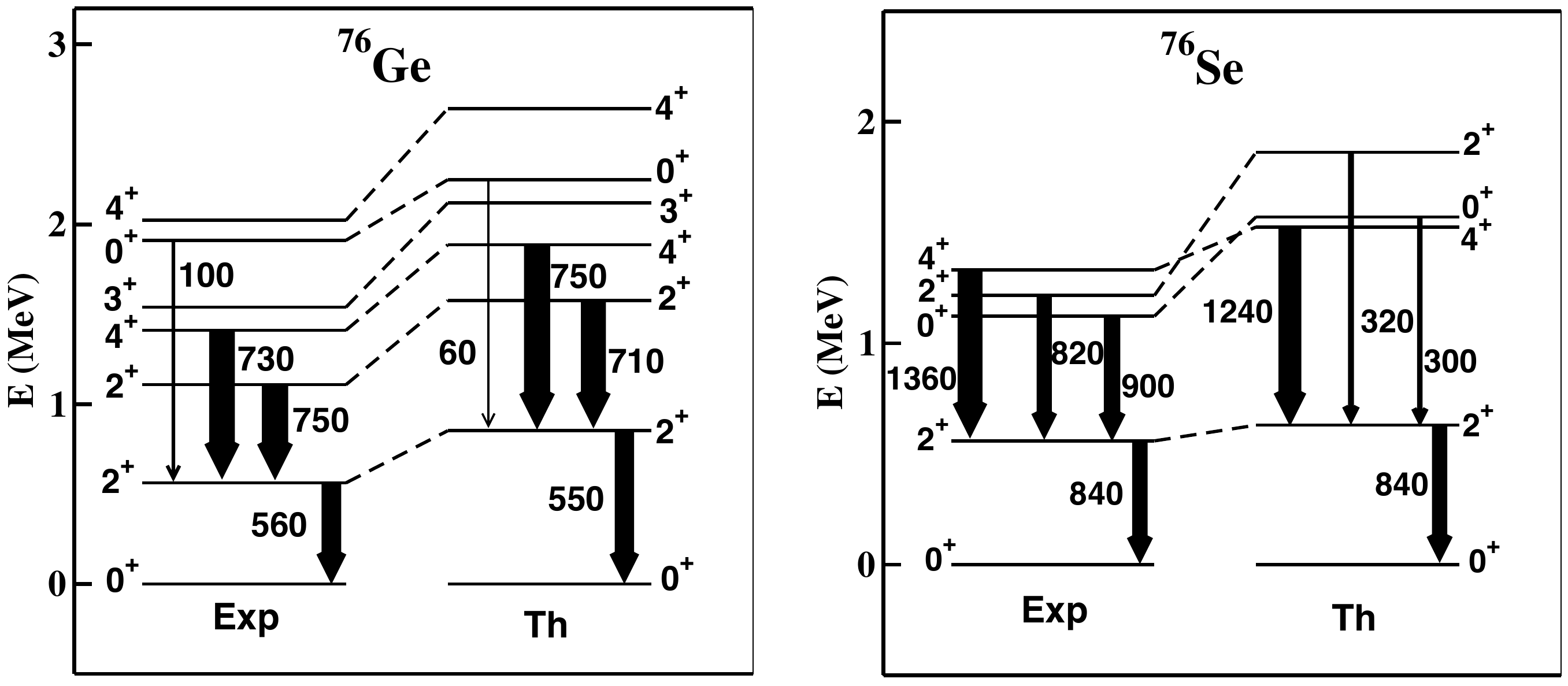}
\caption{ Same as in Fig. \ref{48Ca48Ti}, but for $^{76}$Ge and
  $^{76}$Se (see text for details).}
\label{76Ge76Se}
\end{figure}
\end{center}

The shell-model calculation for $^{76}$Ge and $^{76}$Se are performed
within the model space spanned by the four proton and neutron orbitals
$0f_{5/2}$, $1p_{3/2}$, $1p_{1/2}$ and $0g_{9/2}$, considering
$^{56}$Ni as closed core.
The experimental \cite{ensdf,xundl} and calculated low-energy spectra
of $^{76}$Ge and $^{76}$Se are reported in Fig. \ref{76Ge76Se},
together with the experimental \cite{ensdf,xundl} and
calculated $B(E2)$ strengths (in $e^2{\rm fm}^4$), as in Fig. \ref{48Ca48Ti}

The agreement between the experimental and calculated spectra and
$B(E2)$s is far more satisfactory than that obtained for $A=48$.

In Table \ref{M1_A76} we report the experimental and calculated
$B(M1)$ strengths as well as magnetic dipole moments of $^{76}$Ge and
$^{76}$Se.

\begin{table}[ht]
\caption{ Same as in Table \ref{M1_A48}, but for $^{76}$Ge and
  $^{76}$Se (see text for details).}
\begin{ruledtabular}
\begin{tabular}{cccccc}
\label{M1_A76}
 Nucleus & $J_i \rightarrow J_f $ & $B(M1) _{\rm Expt}$ & I & II & III \\
\colrule
 ~ & ~ & ~ & ~ & ~ & ~\\
 $^{76}$Ge   &          ~                     & ~ & ~ & ~ & ~\\
~  & $2^+_2 \rightarrow 2^+_1$ & $0.003^{+0.002}_{-0.003}$ \cite{Mukhopadhyay17} & 0.006 & 0.004 & 0.005  \\
~  & $4^+_2 \rightarrow 3^+_1$ & $0.02 \pm 0.01$ \cite{Mukhopadhyay17} & 0.062 & 0.027 & 0.025  \\
~ &  ~ & $0.002 \pm 0.001$ \cite{Mukhopadhyay17} & ~ & ~ & ~  \\
~  & $4^+_2 \rightarrow 4^+_1$ & $0.03^{+0.02}_{-0.03}$ \cite{Mukhopadhyay17} & 0.07 & 0.03 & 0.03  \\
~  & ~  & $0.04 \pm 0.02$ \cite{Mukhopadhyay17} & ~ & ~ & ~  \\
 ~ & ~ & ~ & ~ & ~ & ~\\
\colrule
  Nucleus & $J$ & $\mu _{\rm Expt}$ & I & II & III \\
  \colrule
 ~ & ~ & ~ & ~ & ~ & ~\\
 $^{76}$Ge  &     ~        &           ~              &  ~   &  ~   & ~\\
        ~      & $2^+_1$ & $+0.64 \pm 0.02$ \cite{Gurdal13} & +0.53 & +0.83 & +0.84  \\
        ~      & $2^+_2$ & $+0.78 \pm 0.10$ \cite{Gurdal13} & +0.93 & +1.10 & +1.09  \\
        ~      & $4^+_1$ & $+0.96 \pm 0.68$ \cite{Gurdal13} & +0.58 & +1.33 & +1.36  \\
 $^{76}$Se  &     ~        &           ~              &  ~   &  ~   & ~\\
        ~      & $2^+_1$ & $ 0.81 \pm 0.05$ \cite{Speidel98} & +0.37 & +0.60 & +0.58  \\
        ~      & $2^+_2$ & $ 0.70 \pm 0.12$ \cite{Speidel98} & +0.64 & +0.82 & +0.79  \\
        ~      & $4^+_1$ & $ 2.6 \pm 0.4$ \cite{Speidel98} & +0.3 & +0.9 & +0.9  \\
\end{tabular}
\end{ruledtabular}
\end{table}

It can be observed that, with respect to the calculations for
$^{48}$Ca and $^{48}$Ti, now the contribution arising from an
effective transition operator - whose matrix elements are reported in
Table \ref{effM1_56Ni} - is more relevant, and significantly improves
the comparison with data.
This traces back to the fact that, as it is well known
\cite{Towner87}, spin- and spin-isospin-dependent operators need
larger renormalizations when orbitals belonging to the model space
lack their spin-orbit counterpart.
As a matter of fact, this regards single-body matrix elements of the
effective $M1$ - and GT operators - involving the $0f_{5/2}$ and
$0g_{9/2}$ orbitals.
We observe that also for $^{76}$Ge and $^{76}$Se, the blocking effect
on the $M1$ operator seems rather unimportant.

As regards the comparison with experiment, both calculated $B(M1)$s
and dipole moments agree with data, especially those obtained with the
effective operators (II) and (III).
It is worth pointing out that, using the effective operators, the
quenching of the non-diagonal one-body matrix elements in Table
\ref{effM1_56Ni} is responsible for the reduction of the calculated
$B(M1)$s with respect to those obtained with the bare operator.
On the other side, the enhancement of the proton diagonal matrix
element $\langle 0f_{5/2}|| M1 || 0f_{5/2} \rangle$ (see Table
\ref{effM1_56Ni}) leads to an increase of the magnetic dipole moments
of the yrast states.

As can be seen in Table \ref{effGT_56Ni}, the renormalization effect
of the GT operator is even much stronger than that observed for
the $M1$ operator.
This is reflected in our shell-model results for the NMEs of the
$2\nu\beta\beta$ decay of $^{76}$Ge into $^{76}$Se, that are compared
with the experimental value \cite{Barabash15} in Table \ref{ME_76Ge}.

\begin{table}[ht]
  \caption{ Same as in Table \ref{ME_48Ca}, but for the
    $2\nu\beta\beta$ decay of $^{76}$Ge (see text for
    details).}
\begin{ruledtabular}
\begin{tabular}{ccccc}
\label{ME_76Ge}
 Decay & NME$_{\rm Expt}$ & I & II & III \\
\colrule
~ & ~ & ~ & ~ & ~\\
$^{76}$Ge  $\rightarrow$ $^{76}$Se & $ 0.113 \pm 0.006$ & 0.304 & 0.095 & 0.104\\
\colrule
\end{tabular}
\end{ruledtabular}
\end{table}

From the inspection of Table \ref{ME_76Ge}, it can be observed that
using the bare GT operator (I) the calculated NME overestimates the
datum by almost a factor 3, and this gap is recovered employing the
effective operator (II) which introduces and average quenching factor
$q \simeq 0.6$.
As a matter of fact, this renormalization leads to a theoretical
result that is very close to the experimental one.
Moreover, it can be observed a tiny blocking effect that pushes the
calculated value (III) within the experimental error.

\begin{center}
\begin{figure}[ht]
\includegraphics[scale=0.36,angle=0]{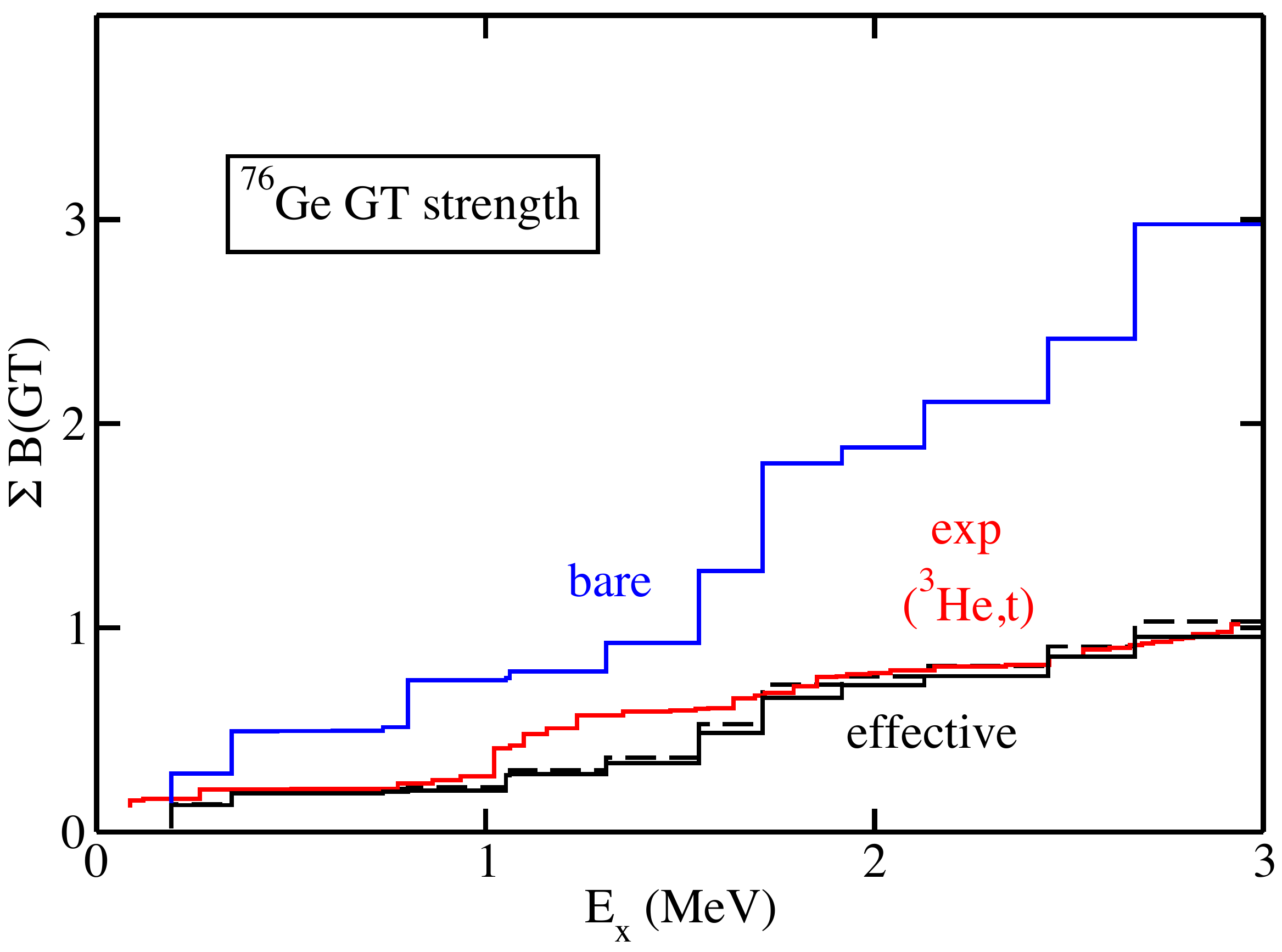}
\caption{Running sums of the $^{76}$Ge $B(p,n)$
  strengths as a function of the excitation energy $E_x$ up to 3 MeV
  (see text for details).}
\label{76GeGT-}
\end{figure}
\end{center}

The role played by the effective operator is also evident when we
compare the calculated and experimental \cite{Thies12a} $\Sigma
B(p,n)$ for $^{76}$Ge as a function of the excitation energy.
This is done in Fig. \ref{76GeGT-}, where  the  running sums of the
GT strengths are reported  up to a 3 MeV excitation energy.
Note that the same notation as in Fig. \ref{48CaGT-} is used here.

This figure, as Table \ref{ME_76Ge}, evidences how crucial it is to take
into account the renormalization of the GT operator to obtain a good
agreement between theory and experiment.
It is worth adding that the contribution of the blocking effect is
almost unrelevant.

For the sake of completeness, we have calculated the theoretical total
${\rm GT}^-$ strengths, and obtained the values 18.2, 6.9, and 7.2
with the bare operator (I), and the effective ones (II) and (III), respectively.

\subsection{$^{82}$Se}\label{82Se}
\begin{center}
\begin{figure}[ht]
\includegraphics[scale=0.32,angle=0]{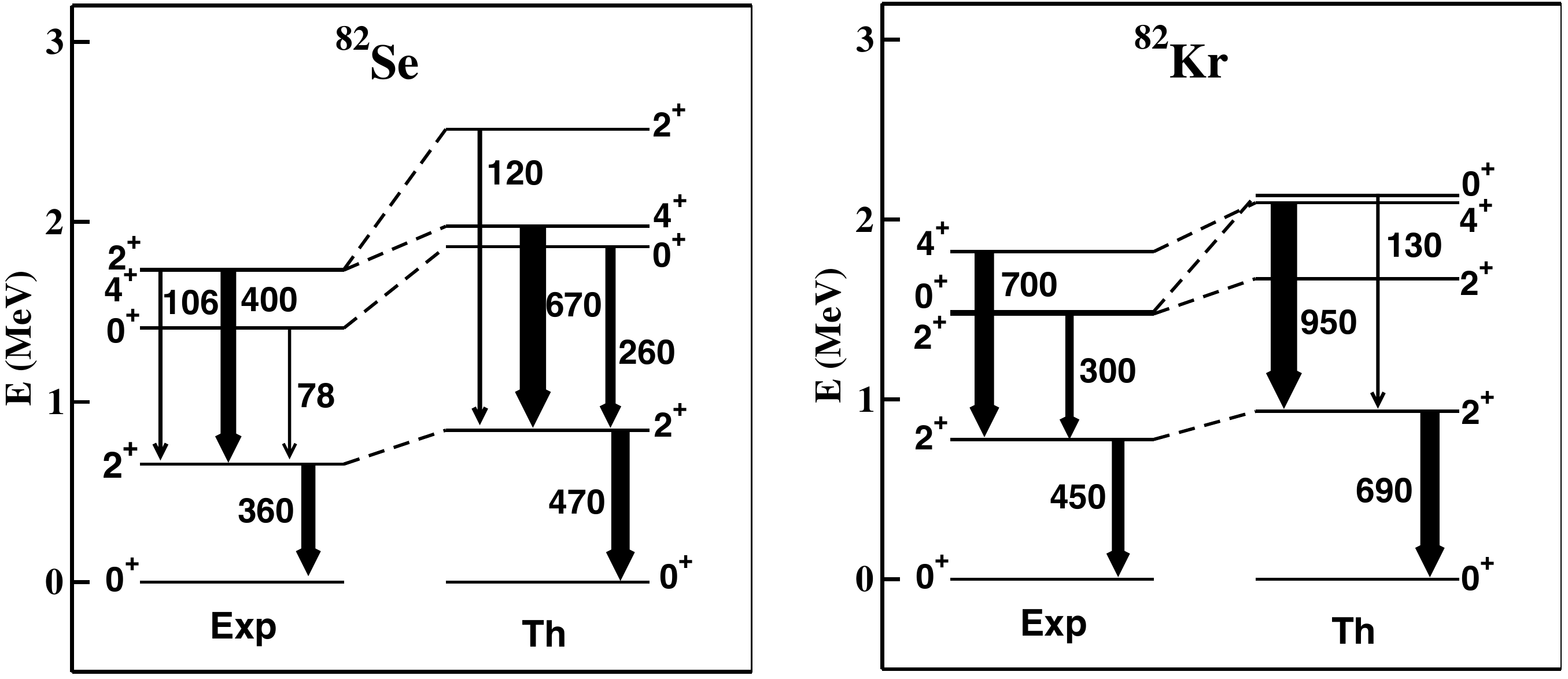}
\caption{ Same as in Fig. \ref{48Ca48Ti}, but for $^{82}$Se and
  $^{82}$Kr (see text for details).}
\label{82Se82Kr}
\end{figure}
\end{center}

As for $^{76}$Ge and $^{76}$Se, the shell model calculation for
$^{82}$Se and $^{82}$Kr has been carried out using, as model space,
the four proton and neutron orbitals $0f_{5/2}$, $1p_{3/2}$,$1p_{1/2}$
and $0g_{9/2}$ placed outside  $^{56}$Ni.
In Fig. \ref{82Se82Kr} the calculated low-energy spectra and $B(E2)$s
are compared with experiment \cite{ensdf,xundl}.

The agreement between theory and experiment can be considered
satisfactory, the largest discrepancy, in both nuclei, occurring for
the $B(E2; 0^+_2\rightarrow  2^+_1)$s, whose calculated values are
about a factor 3 smaller than the observed ones.

As regards the observables linked to the $M1$ operator, the only
available data for the low-lying states reported in
Fig. \ref{82Se82Kr} are the magnetic dipole moments shown in Table
\ref{M1_A82}.

\begin{table}[ht]
\caption{Experimental and calculated magnetic dipole moments (in ${\rm
    nm }$) of $^{82}$Se and $^{82}$Kr. We report those for the
  observed states in Fig. \ref{82Se82Kr}.}
\begin{ruledtabular}
\begin{tabular}{cccccc}
\label{M1_A82}
 Nucleus & $J$ & $\mu _{\rm Expt}$ & I & II & III \\
\colrule
 ~ & ~ & ~ & ~ & ~ & ~\\
 $^{82}$Se  &     ~        &           ~              &  ~   &  ~   & ~\\
        ~      & $2^+_1$ & $+0.99 \pm 0.06$ \cite{ensdf} & +0.72 & +1.03 & +1.05  \\
        ~      & $4^+_1$ & $2.3 \pm 1.5$ \cite{ensdf} & +1.17 & +1.88 & +1.93  \\
 $^{82}$Kr  &     ~        &           ~              &  ~   &  ~   & ~\\
        ~      & $2^+_1$ & $ +0.80 \pm 0.04$ \cite{ensdf} & +0.50 & +0.83 & +0.83  \\
        ~      & $4^+_1$ & $ +1.2 \pm 0.8$ \cite{ensdf} & +0.5 & +1.3 & +1.3  \\
\end{tabular}
\end{ruledtabular}
\end{table}

Since the model space is the same as for $A=76$ nuclei, the matrix
elements of the effective $M1$ operator (II) are those reported in
Table \ref{effM1_56Ni}.
The action of the effective operators, as can be observed from the
inspection of Table \ref{M1_A82}, is to improve the comparison with the
data of the shell-model results, with respect to those obtained with
the bare operator (I).
This result evidences the role of the renormalization of the bare
operator to take into account the degrees of freedom that have been
left out by constraining the nuclear wave function to the valence
nucleons interacting in the truncated model space.

As for the $2\nu\beta\beta$ decay of $^{76}$Ge, the quenching of the
matrix elements of the GT operator, shown in Table \ref{effGT_56Ni},
is crucial to improve our calculation of the NME of the decay of
$^{82}$Se into $^{82}$Kr.
In fact, the NME calculated with the bare operator overestimates the
experimental value \cite{Barabash15} by a factor 4, as can be
inferred from Table \ref{ME_82Se}, while calculations performed with
the effective operators (II-III) provide far better results.

\begin{table}[ht]
\caption{Same as in Table \ref{ME_48Ca}, but for the $2\nu\beta\beta$
  decay of $^{82}$Se (see text for details).}
\begin{ruledtabular}
\begin{tabular}{ccccc}
\label{ME_82Se}
 Decay & NME$_{\rm Expt}$ & I & II & III \\
\colrule
~ & ~ & ~ & ~ & ~\\
$^{82}$Se  $\rightarrow$ $^{82}$Kr & $ 0.083 \pm 0.004$ & 0.347 & 0.111 & 0.109 \\
\colrule
\end{tabular}
\end{ruledtabular}
\end{table}

\begin{center}
\begin{figure}[ht]
\includegraphics[scale=0.36,angle=0]{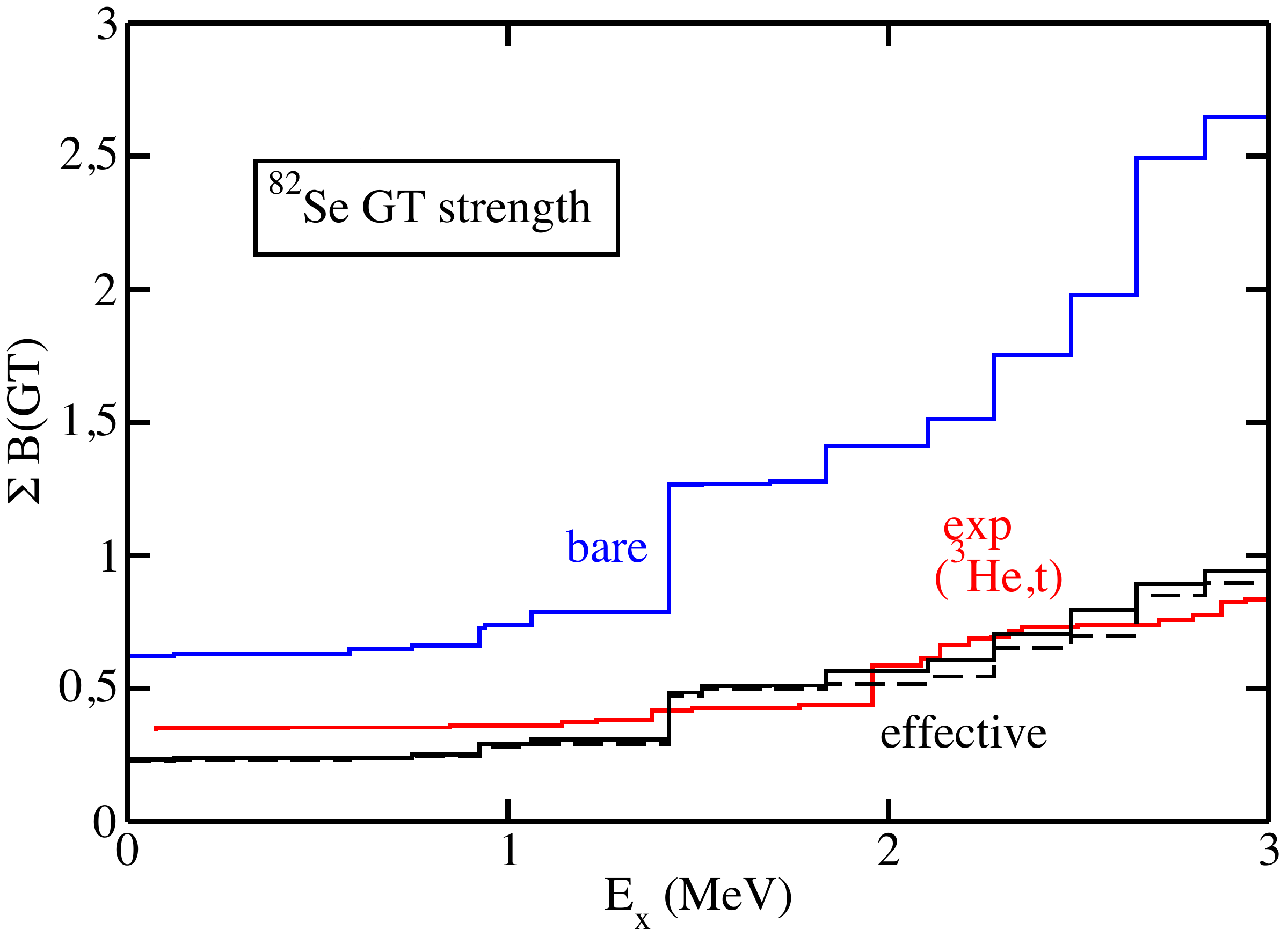}
\caption{Running sums of the $^{82}$Se $B(p,n)$
  strengths as a function of the excitation energy $E_x$ up to 3 MeV
  (see text for details).}
\label{82SeGT-}
\end{figure}
\end{center}

The quenching of the effective GT operator is a feature that is
crucial also to provide a calculated $\Sigma B(p,n)$ curve for
$^{82}$Se, as a function of the excitation energy, that almost
overlaps with the experimental one \cite{Frekers16}, as can be seen in
Fig. \ref{82SeGT-} where the running sums of the $^{82}$Se GT
strengths up to a 3 MeV excitation energy are reported.

As for the calculations of $^{48}$Ca, $^{76}$Ge $\Sigma B(p,n)$, we
observe a negligible role of the blocking effect.

We conclude this section reporting the calculated total ${\rm GT}^-$
strengths that are 21.6, 8.5, and 8.9 with the bare operator (I), and the
effective ones (II) and (III), respectively.

\subsection{$^{130}$Te}\label{130Te}
\begin{center}
\begin{figure}[ht]
\includegraphics[scale=0.30,angle=0]{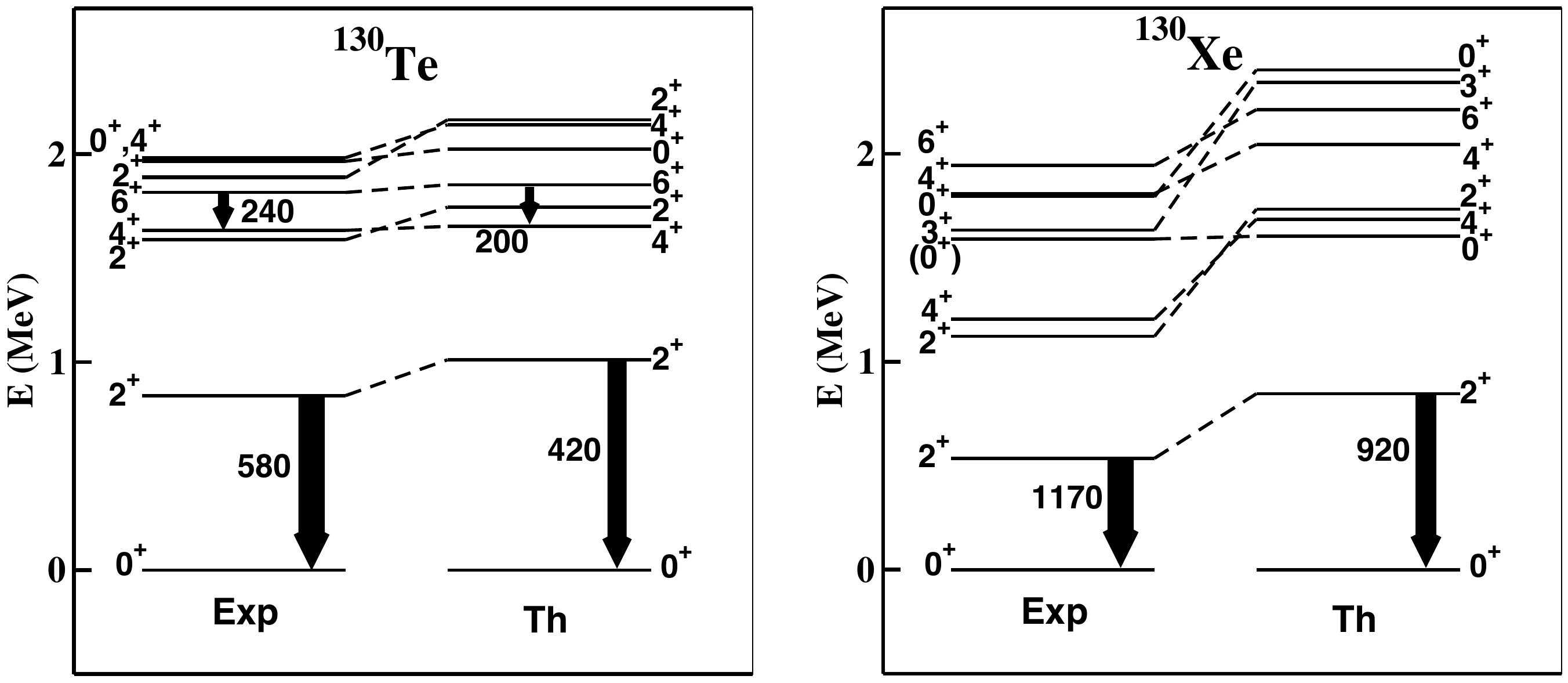}
\caption{ Same as in Fig. \ref{48Ca48Ti}, but for $^{130}$Te and
  $^{130}$Xe (see text for details).}
\label{130Te130Xe}
\end{figure}
\end{center}

The shell-model calculation for $^{130}$Te and $^{130}$Xe are
performed within the model space spanned by the five proton and
neutron orbitals $0g_{7/2}$, $1d_{5/2}$,$1d_{3/2}$, $2s_{1/2}$ and
$0h_{11/2}$, considering  $^{100}$Sn as closed core.
For the sake of completeness, the experimental \cite{ensdf,xundl} and
calculated low-energy spectra and $B(E2)$s, already reported
in Ref. \cite{Coraggio17a}, are also presented in this work in
Fig. \ref{130Te130Xe}.

From inspection of Fig. \ref{130Te130Xe}, we observe that the
comparison between the calculated and experimental low-energy spectra
is very good for $^{130}$Te, while is less satisfactory for $^{130}$Xe.
As regards the calculated $B(E2)$s, they compare well with the observed
values for both nuclei, providing good expectations about the reliability of the SM
wavefunctions.

In Table \ref{M1_A130} the calculated $B(M1;2^+_3 \rightarrow 2^+_1)$
of $^{130}$Te is reported and compared with the two experimental values of Ref. \cite{Hicks08}.
In the same table the calculated and observed magnetic dipole moments
of $^{130}$Te and $^{130}$Xe can be found.

\begin{table}[ht]
\caption{ Same as in Table \ref{M1_A48}, but for $^{130}$Te and
  $^{130}$Xe (see text for details). We report those for the
  observed states in Fig. \ref{130Te130Xe}.}
\begin{ruledtabular}
\begin{tabular}{cccccc}
\label{M1_A130}
 Nucleus & $J_i \rightarrow J_f $ & $B(M1) _{\rm Expt}$ & I & II & III \\
\colrule
 ~ & ~ & ~ & ~ & ~ & ~\\
 $^{130}$Te   &          ~                     & ~ & ~ & ~ & ~\\
~  & $2^+_3 \rightarrow 2^+_1$ & $0.037^{+0.03}_{-0.04}$ \cite{Hicks08} & 0.057 & 0.085 & 0.077  \\
~  & ~  & $0.097^{+0.08}_{-0.11}$ \cite{Hicks08} & ~ & ~ & ~  \\
\colrule
 Nucleus & $J$ & $\mu _{\rm Expt}$ & I & II & III \\
\colrule
 ~ & ~ & ~ & ~ & ~ & ~\\
 $^{130}$Te  &     ~        &           ~              &  ~   &  ~   & ~\\
        ~      & $2^+_1$ & $0.58 \pm 0.10$ \cite{ensdf} & +0.52 & +0.71 & +0.71  \\
 $^{130}$Xe  &     ~        &           ~              &  ~   &  ~   & ~\\
        ~      & $2^+_1$ & $ 0.57 \pm 0.14$ \cite{ensdf} & +0.50 & +0.67 & +0.66  \\
\end{tabular}
\end{ruledtabular}
\end{table}

As can be seen, similarly to the results of the calculations for
$^{76}$Ge and $^{82}$Se, the role of the effective $M1$ operator is
relevant.
As a matter of fact, the smaller $B(M1)$ values, compared with the one
calculated with the bare operator, are a consequence of the general
quenching of the non-diagonal matrix elements reported in Table
\ref{effM1_100Sn}.
On the other side, the enhancement of the proton $0g_{7/2}$ diagonal
matrix element is responsible for the larger dipole moments, when they
are calculated employing the effective operators (II) and (III).
Actually, because of the large experimental errors, it is not clear if
the effective operators are able to provide a better agreement with
experiment for the dipole moments with respect to the bare operator.
As regards the $B(M1;2^+_3 \rightarrow 2^+_1)$, it turns out that our
calculated values are closer to the smallest of the two values
reported in Ref. \cite{Hicks08}.
Finally, it is worth noting that there is no sizeable role of the
blocking effect.

The calculated and experimental values of the NME for the $^{130}$Te
$2\nu\beta\beta$ decay \cite{Barabash15} are reported in Table \ref{ME_130Te}.

\begin{table}[ht]
\caption{Same as in Table \ref{ME_48Ca}, but for the
    $2\nu\beta\beta$ decay (in MeV$^{-1}$) of $^{130}$Te (see text for
    details).}
\begin{ruledtabular}
\begin{tabular}{ccccc}
\label{ME_130Te}
 Decay & NME$_{\rm Expt}$ & I & II & III \\
\colrule
~ & ~ & ~ & ~ & ~\\
$^{130}$Te  $\rightarrow$ $^{130}$Xe & $ 0.031 \pm 0.004$ & 0.131 &
                                                                    0.057 & 0.061 \\
\colrule
\end{tabular}
\end{ruledtabular}
\end{table}

As shown in our previous study \cite{Coraggio17a}, where the matrix
elements of the effective GT operator can be found in Tables
(III-IV), the quenching of the bare operator (I) provided by the
effective ones (II-III) plays a fundamental role to obtain a
reasonable comparison with the experimental NME.
As a matter of fact, our shell model calculation gives a
$2\nu\beta\beta$ NME that is almost 4 times bigger than the
experimental one, starting from GT operator (I).
On the other hand, the effective operators, derived via many-body
perturbation theory, take into account the reduction of the full
Hilbert space to configurations constrained by the valence nucleons
interacting in the model space and provide NMEs that are almost within
experimental error bars.

\begin{center}
\begin{figure}[ht]
\includegraphics[scale=0.36,angle=0]{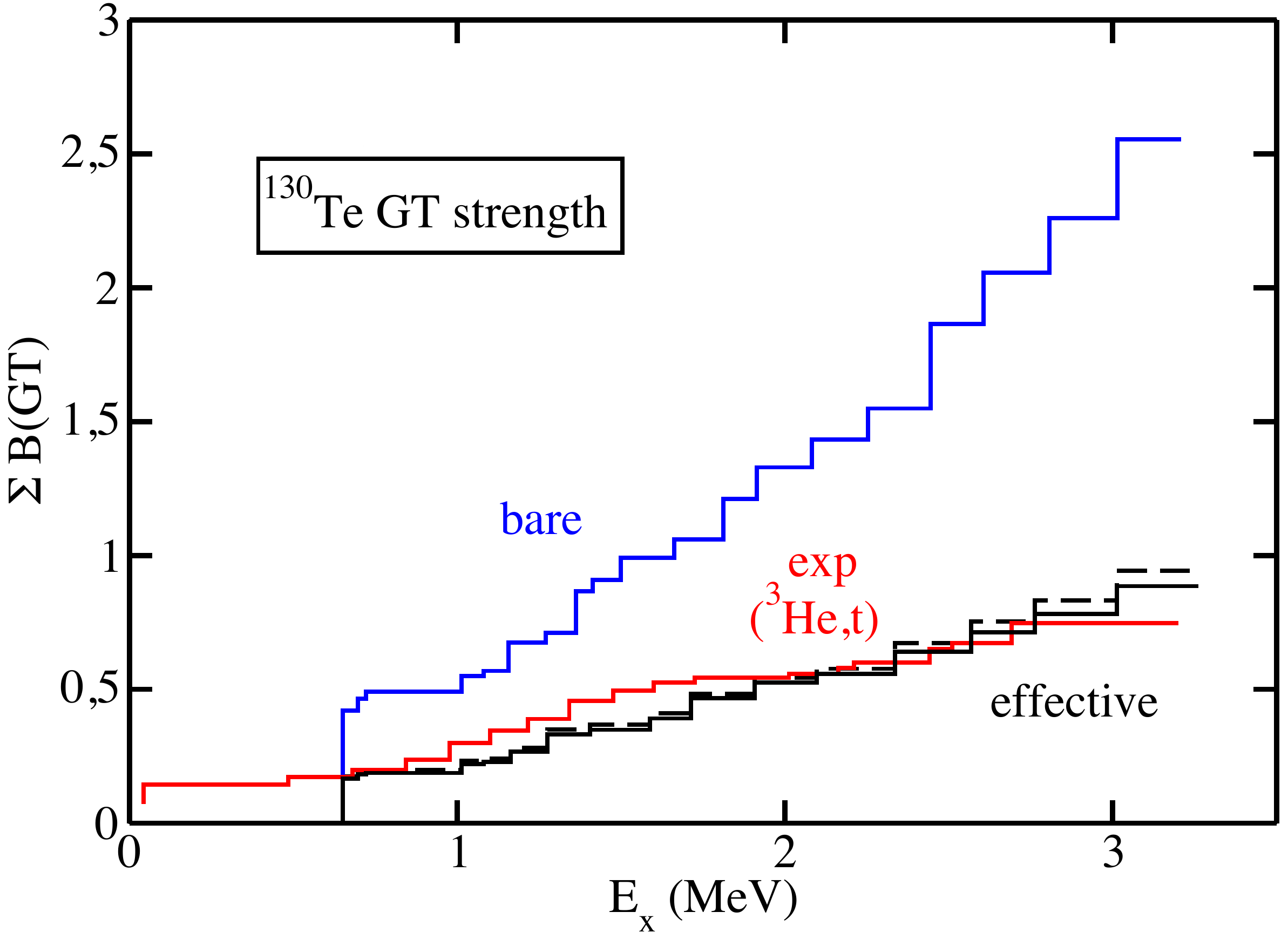}
\caption{Running sums of the $^{130}$Te $B(p,n)$
  strengths as a function of the excitation energy $E_x$ up to 3 MeV
  (see text for details).}
\label{130TeGT-}
\end{figure}
\end{center}

These considerations hold, obvioulsy, also for the calculation of the
$^{130}$Te $\Sigma B(p,n)$, whose results are reported in
Fig. \ref{130TeGT-} and compared with available data \cite{Puppe12} up
to 3 MeV excitation energy.

As for the $^{76}$Ge and $^{82}$Se running sums, the curves obtained
with the effective operators (II-III) lie much closer to the
experimental one than that calculated employing the bare operator
(I), and almost overlap each other.

The total ${\rm GT}^-$ strengths, obtained with effective operators
(I-III), are 46.4, 18.3, and 18.6, respectively.

\subsection{$^{136}$Xe}\label{136Xe}
The shell-model calculations for $^{136}$Xe and $^{136}$Ba are carried
out using the same model space, effective Hamiltonian and transition
operators as for $^{130}$Te and $^{130}$Xe, and details about SP
energies, TBMEs, effective charges, and effective GT matrix elements
can be found in Ref. \cite{Coraggio17a}.

\begin{center}
\begin{figure}[ht]
\includegraphics[scale=0.30,angle=0]{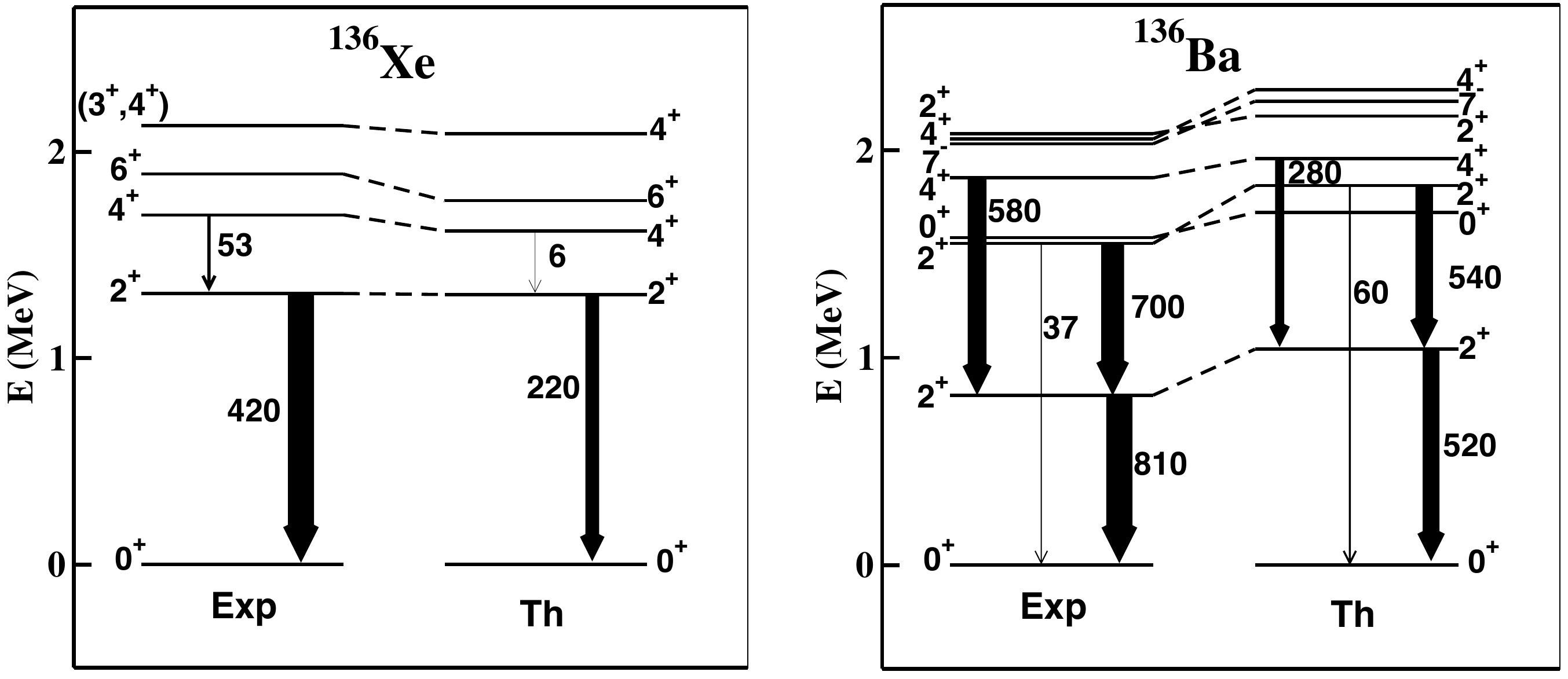}
\caption{ Same as in Fig. \ref{48Ca48Ti}, but for $^{136}$Xe and
  $^{136}$Ba (see text for details).}
\label{136Xe136Ba}
\end{figure}
\end{center}

We present, as in our previous study, the experimental
\cite{ensdf,xundl} and calculated low-energy spectra and $B(E2)$s
which we have reported in Fig. \ref{136Xe136Ba}.

The comparison between theory and experiment, as regards the low-lying
excited states and the $B(E2)$ transition rates, is excellent for both
nuclei, testifying once more the reliability of the realistic shell model.

In Table \ref{M1_A136} we report the calculated and experimental
$B(M1)$s of $^{136}$Ba, involving some of the excited states
reported in Fig. \ref{136Xe136Ba}, together with the $J=2^+_1$ magnetic
dipole moment.
We compare also theory and experiment for the $J=2^+_1,4^+_1$ magnetic
dipole moments of $^{136}$Xe.

\begin{table}[ht]
\caption{ Same as in Table \ref{M1_A48}, but for $^{136}$Xe and
  $^{136}$Ba (see text for details). We report those for the
  observed states in Fig. \ref{136Xe136Ba}.}
\begin{ruledtabular}
\begin{tabular}{cccccc}
\label{M1_A136}
 Nucleus & $J_i \rightarrow J_f $ & $B(M1) _{\rm Expt}$ & I & II & III \\
\colrule
 ~ & ~ & ~ & ~ & ~ & ~\\
 $^{136}$Ba   &          ~                     & ~ & ~ & ~ & ~\\
~  & $2^+_2 \rightarrow 2^+_1$ & $0.02 \pm 0.1$ \cite{Mukhopadhyay08} & 0.07 & 0.06 & 0.06  \\
~  & $2^+_3 \rightarrow 2^+_1$ & $0.002 \pm 0.002$ \cite{Mukhopadhyay08} & 0.006 & 0.002 & 0.001  \\
~  & $4^+_2 \rightarrow 4^+_1$ & $0.06^{+0.08}_{-0.05}$ \cite{Mukhopadhyay08} & 0.15 & 0.10 & 0.09  \\
 ~ & ~ & ~ & ~ & ~ & ~\\
\colrule
 Nucleus & $J$ & $\mu _{\rm Expt}$ & I & II & III \\
\colrule
 ~ & ~ & ~ & ~ & ~ & ~\\
 $^{136}$Xe  &     ~        &           ~              &  ~   &  ~   & ~\\
        ~      & $2^+_1$ & $1.53 \pm 0.09$ \cite{ensdf} & +1.05 & +1.15 & +1.14  \\
        ~      & $4^+_1$ & $3.2 \pm 0.6$ \cite{ensdf} & +2.02 & +2.24 & +2.22  \\
 $^{136}$Ba  &     ~        &           ~              &  ~   &  ~   & ~\\
        ~      & $2^+_1$ & $ 0.69 \pm 0.10$ \cite{ensdf} & +0.48 & +0.60 & +0.59  \\
\end{tabular}
\end{ruledtabular}
\end{table}

As a matter of fact, we observe the same tendency we have found in the
previous calculations, that is the quenching of $B(M1)$ values
obtained with effective operators (II-III), and the enhancement of the
dipole moments when the same operators are employed.
This is grounded on the same observations we have made in
Section \ref{130Te}, and supported by the inspection of the list of
the matrix elements in Table \ref{effM1_100Sn}.

Actually, both features lead to an improvemement in the description of
the data, and support again the crucial role of the renormalization of
transition operators by way of the many-body perturbation theory.

This consideration is even more valid when we consider the calculation
of the NME for the $^{136}$Xe $2\nu\beta\beta$ decay, whose results
are reported in Table \ref{ME_136Xe} and compared with the datum
\cite{Barabash15}.

\begin{table}[ht]
\caption{ Same as in Table \ref{ME_48Ca}, but for the $2\nu\beta\beta$
  decay (in MeV$^{-1}$) of $^{136}$Xe (see text for details).}
\begin{ruledtabular}
\begin{tabular}{ccccc}
\label{ME_136Xe}
 Decay & NME$_{\rm Expt}$ & I & II & III \\
\colrule
~ & ~ & ~ & ~ & ~\\
$^{136}$Xe  $\rightarrow$ $^{136}$Ba & $ 0.0181 \pm 0.0007$ & 0.0910 & 0.0332 & 0.0341\\
\colrule
\end{tabular}
\end{ruledtabular}
\end{table}

\begin{center}
\begin{figure}[ht]
\includegraphics[scale=0.36,angle=0]{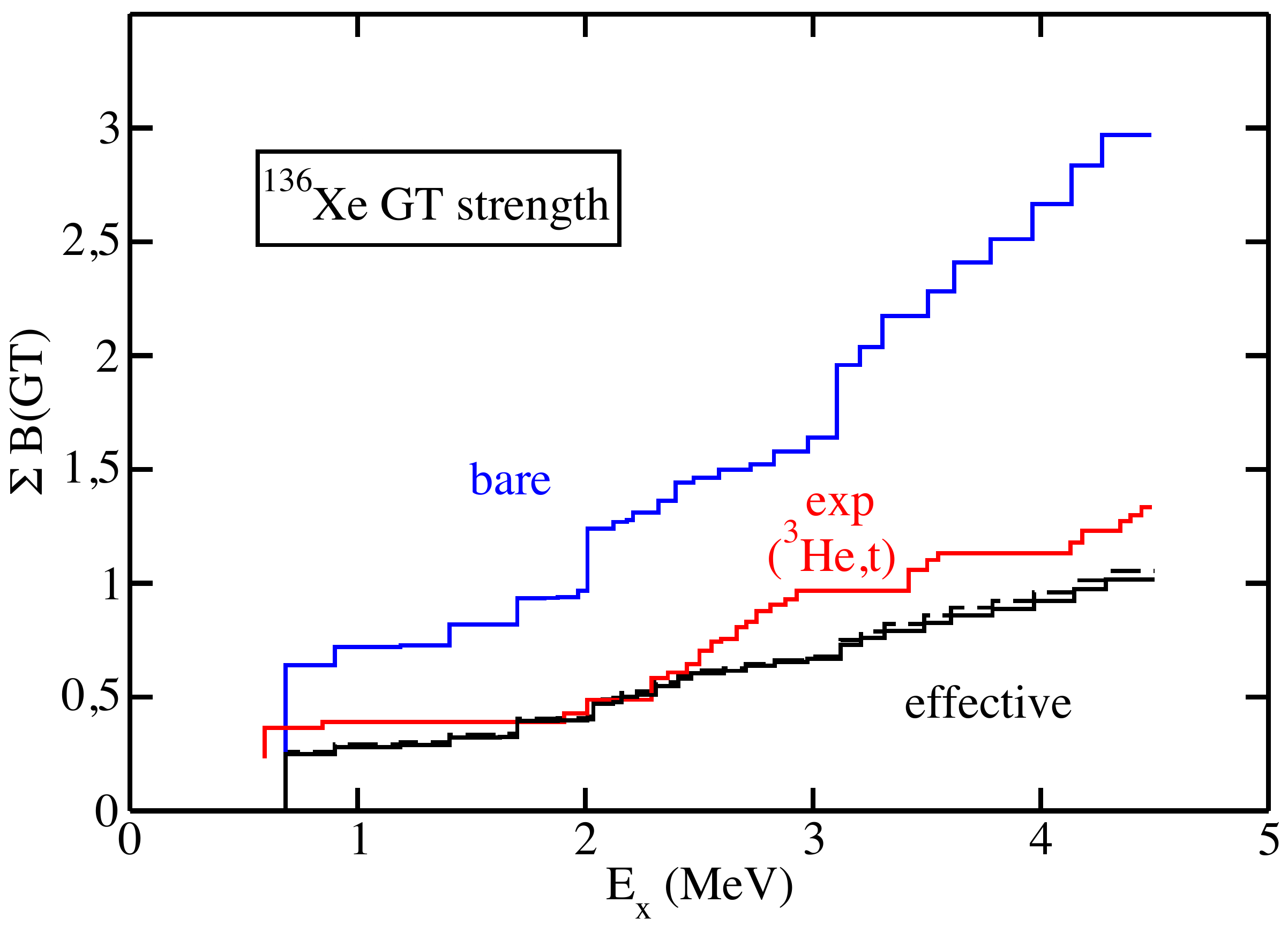}
\caption{Running sums of the $^{136}$Xe $B(p,n)$
  strengths as a function of the excitation energy $E_x$ up to 4.5 MeV
  (see text for details).}
\label{136XeGT-}
\end{figure}
\end{center}

We see that the (II-III) NMEs are more than a factor 3 smaller than
the value obtained with the bare operator (I), and closer to the
experimental value.
The same feature comes out in Fig. \ref{136XeGT-}, where we report the
calculated and experimental \cite{Frekers13} $\Sigma B(p,n)$ of
$^{136}$Xe up to 4.5 MeV excitation energy.

Also for the $^{136}$Xe running sums, the many-body renormalization of
the GT operator is crucial to reproduce the experimental curve, with a
negligible contribution of the blocking effect.

The total ${\rm GT}^-$ strengths, obtained with bare and effective
operators (I-III), are 51.9, 20.7, and 21.0, respectively.

\section{Conclusions and Outlook}\label{conclusions}

In this paper we have studied the role of effective operators to
calculate, within the realistic shell model, observables that are
related to spin- and spin-isospin-dependent transitions.
Our main focus has been on GT transitions for nuclei that are
candidates for the detection of the $0 \nu\beta\beta$-decay, and we
have calculated, for several nuclei and over a wide mass range,
$2\nu\beta\beta$-decay NMEs and the running sums of the $B(p,n)$
strengths to compare them with the available data.
Since the magnetic-dipole $M1$ operator incorporates an isovector-spin
term with the same structure of the GT operator, we have extended this
analysis to the calculation of $B(M1)$s and magnetic dipole moments to
strengthen our investigation.

As a matter of fact, our aim has been to demonstrate that the present
status of the many-body perturbation theory allows to derive
consistently effective Hamiltonians and transition operators that are
able to reproduce quantitatively the observed spectroscopic and decay
properties, without resorting to an empirical quenching of the axial
coupling constant $g_A$, or to empirically fitted spin and orbital
$g$-factors $g_s,g_l$.

The quenching factors corresponding to the matrix eIements of the
effective $M1$ and GT operators are reported in Tables
\ref{effM1_40Ca}-\ref{effM1_100Sn} and Tables
\ref{effGT_40Ca}-\ref{effGT_100Sn}, respectively.
It is worth noting  that the calculated quenching effect on the $M1$
operator is overall smaller than for GT transitions, which points to
the fact that the two operators are differently affected by the
renormalization procedure. 
This result highlights that for the renormalization of the $M1$
operator a non-negligible role is played by its isoscalar and
isovector orbital components.
As a matter of fact, from the inspection of these tables, the
quenching of proton-proton $M1$ matrix elements is overall largely
different from the $GT$ one, the latter being much closer to that
obtained for neutron-neutron $M1$ matrix elements (which own the spin
component only).

\begin{center}
\begin{figure}[ht]
\includegraphics[scale=0.44,angle=0]{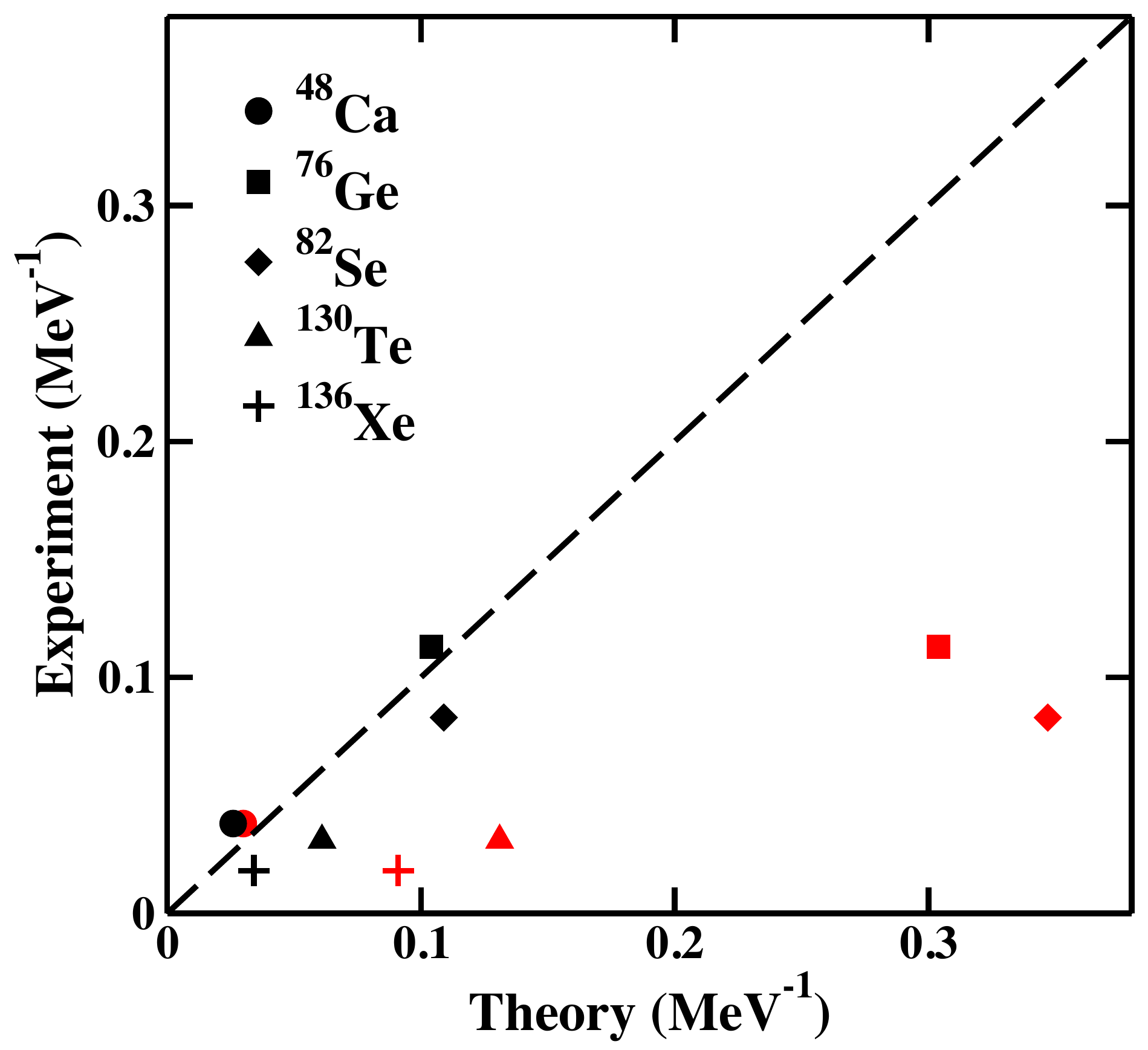}
\caption{Correlation plot between the calculated (x-axis) and the
  experimental (y-axis) $2\nu\beta\beta$ decay NMEs (see text for
  details).}
\label{correlation}
\end{figure}
\end{center}

In order to show and stress pictorially the main outcome of our study
about the relevance played by effective transition operators, in
Fig. \ref{correlation} we report a correlation plot between our
calculated $2\nu\beta\beta$ decay NMEs and the corresponding
experimental values.
The quantities in Fig. \ref{correlation} are already reported in Tables
\ref{ME_48Ca},\ref{ME_76Ge},\ref{ME_82Se},\ref{ME_130Te},\ref{ME_136Xe}.

The red symbols correspond to the results obtained employing the bare
operators (I), while the black ones indicate the results obtained with
the effective operators (III).

As can be seen, the red points are all spread on the lower side of
the figure, except the one corresponding to $^{48}$Ca, and lie far
away from the identity, that is represented by a dashed line.
This feature characterizes the nuclei that are described by way of a
model space where some of its orbitals lack their spin-orbit
counterparts, leading to an overestimation of the calculated NME with
respect to the experimental value.

The black points, that correspond to the effective GT operators, on
the other hand regroup themselves close to the identity, as a reliable
calculation should do.

It is worth reminding that our results may be traced back to earlier
investigations carried out by Towner and collaborators since the 1980s
(see for instance \cite{Towner83,Towner87,Johnstone98,Brown05}), where
the role of microscopically derived effective spin-dependent operators is
enlightened. 
Present work takes advantage of modern developments to derive the
effective shell-model Hamiltonians and operators (see for example
Refs. \cite{Coraggio12a,Suzuki95}), and up-to-date approaches to the
renormalization of realistic $NN$ potentials \cite{Bogner02}.

On the above grounds, we intend to extend our study by investigating
the role of meson-exchange corrections to the electroweak currents
\cite{Park93,Pastore09,Piarulli13,Baroni16b}.
More precisely, we aim in a near future at building up effective
shell-model Hamiltonians and operators starting from two- and
three-body nuclear potentials derived within the framework of chiral
perturbation theory \cite{Coraggio18a}, and taking also into account
the contributions of chiral two-body electroweak currents to the
effective GT operators.
As a matter of fact, recent studies have shown that $\beta$- and
neutrinoless double-beta decays may be significantly affected by these
contributions \cite{Menendez11,Wang18}, when consistently starting
from chiral potentials.

At last, our final goal is to benefit from the expertise we have
gained to evaluate the $0 \nu \beta \beta$ decay NMEs for the nuclei
studied in present paper \cite{Coraggio19b}.
 
\bibliographystyle{apsrev}
\bibliography{biblio.bib}

\begin{thebibliography}{80}
\expandafter\ifx\csname natexlab\endcsname\relax\def\natexlab#1{#1}\fi
\expandafter\ifx\csname bibnamefont\endcsname\relax
  \def\bibnamefont#1{#1}\fi
\expandafter\ifx\csname bibfnamefont\endcsname\relax
  \def\bibfnamefont#1{#1}\fi
\expandafter\ifx\csname citenamefont\endcsname\relax
  \def\citenamefont#1{#1}\fi
\expandafter\ifx\csname url\endcsname\relax
  \def\url#1{\texttt{#1}}\fi
\expandafter\ifx\csname urlprefix\endcsname\relax\def\urlprefix{URL }\fi
\providecommand{\bibinfo}[2]{#2}
\providecommand{\eprint}[2][]{\url{#2}}

\bibitem[{\citenamefont{Hjorth-Jensen et~al.}(2017)\citenamefont{Hjorth-Jensen,
  Lombardo, and van Kolck}}]{HjorthJensen17}
\bibinfo{editor}{\bibfnamefont{M.}~\bibnamefont{Hjorth-Jensen}},
  \bibinfo{editor}{\bibfnamefont{M.~P.} \bibnamefont{Lombardo}},
  \bibnamefont{and} \bibinfo{editor}{\bibfnamefont{U.}~\bibnamefont{van
  Kolck}}, eds., \emph{\bibinfo{title}{Lecture Notes in Physics}}, vol.
  \bibinfo{volume}{936} (\bibinfo{publisher}{Springer}, \bibinfo{year}{2017}).

\bibitem[{\citenamefont{Hjorth-Jensen et~al.}(1995)\citenamefont{Hjorth-Jensen,
  Kuo, and Osnes}}]{Hjorth95}
\bibinfo{author}{\bibfnamefont{M.}~\bibnamefont{Hjorth-Jensen}},
  \bibinfo{author}{\bibfnamefont{T.~T.~S.} \bibnamefont{Kuo}},
  \bibnamefont{and} \bibinfo{author}{\bibfnamefont{E.}~\bibnamefont{Osnes}},
  \bibinfo{journal}{Phys. Rep.} \textbf{\bibinfo{volume}{261}},
  \bibinfo{pages}{125} (\bibinfo{year}{1995}).

\bibitem[{\citenamefont{Coraggio et~al.}(2012)\citenamefont{Coraggio, Covello,
  Gargano, Itaco, and Kuo}}]{Coraggio12a}
\bibinfo{author}{\bibfnamefont{L.}~\bibnamefont{Coraggio}},
  \bibinfo{author}{\bibfnamefont{A.}~\bibnamefont{Covello}},
  \bibinfo{author}{\bibfnamefont{A.}~\bibnamefont{Gargano}},
  \bibinfo{author}{\bibfnamefont{N.}~\bibnamefont{Itaco}}, \bibnamefont{and}
  \bibinfo{author}{\bibfnamefont{T.~T.~S.} \bibnamefont{Kuo}},
  \bibinfo{journal}{Ann. Phys.} \textbf{\bibinfo{volume}{327}},
  \bibinfo{pages}{2125} (\bibinfo{year}{2012}).

\bibitem[{\citenamefont{Coraggio et~al.}(2009)\citenamefont{Coraggio, Covello,
  Gargano, Itaco, and Kuo}}]{Coraggio09a}
\bibinfo{author}{\bibfnamefont{L.}~\bibnamefont{Coraggio}},
  \bibinfo{author}{\bibfnamefont{A.}~\bibnamefont{Covello}},
  \bibinfo{author}{\bibfnamefont{A.}~\bibnamefont{Gargano}},
  \bibinfo{author}{\bibfnamefont{N.}~\bibnamefont{Itaco}}, \bibnamefont{and}
  \bibinfo{author}{\bibfnamefont{T.~T.~S.} \bibnamefont{Kuo}},
  \bibinfo{journal}{Prog. Part. Nucl. Phys.} \textbf{\bibinfo{volume}{62}},
  \bibinfo{pages}{135} (\bibinfo{year}{2009}).

\bibitem[{\citenamefont{Ellis and Osnes}(1977)}]{Ellis77}
\bibinfo{author}{\bibfnamefont{P.~J.} \bibnamefont{Ellis}} \bibnamefont{and}
  \bibinfo{author}{\bibfnamefont{E.}~\bibnamefont{Osnes}},
  \bibinfo{journal}{Rev. Mod. Phys.} \textbf{\bibinfo{volume}{49}},
  \bibinfo{pages}{777} (\bibinfo{year}{1977}).

\bibitem[{\citenamefont{Suzuki and Okamoto}(1995)}]{Suzuki95}
\bibinfo{author}{\bibfnamefont{K.}~\bibnamefont{Suzuki}} \bibnamefont{and}
  \bibinfo{author}{\bibfnamefont{R.}~\bibnamefont{Okamoto}},
  \bibinfo{journal}{Prog. Theor. Phys.} \textbf{\bibinfo{volume}{93}},
  \bibinfo{pages}{905} (\bibinfo{year}{1995}).

\bibitem[{\citenamefont{Siiskonen et~al.}(2001)\citenamefont{Siiskonen,
  Hjorth-Jensen, and Suhonen}}]{Siiskonen01}
\bibinfo{author}{\bibfnamefont{T.}~\bibnamefont{Siiskonen}},
  \bibinfo{author}{\bibfnamefont{M.}~\bibnamefont{Hjorth-Jensen}},
  \bibnamefont{and} \bibinfo{author}{\bibfnamefont{J.}~\bibnamefont{Suhonen}},
  \bibinfo{journal}{Phys. Rev. C} \textbf{\bibinfo{volume}{63}},
  \bibinfo{pages}{055501} (\bibinfo{year}{2001}).

\bibitem[{\citenamefont{Holt and Engel}(2013)}]{Holt13d}
\bibinfo{author}{\bibfnamefont{J.~D.} \bibnamefont{Holt}} \bibnamefont{and}
  \bibinfo{author}{\bibfnamefont{J.}~\bibnamefont{Engel}},
  \bibinfo{journal}{Phys. Rev. C} \textbf{\bibinfo{volume}{87}},
  \bibinfo{pages}{064315} (\bibinfo{year}{2013}).

\bibitem[{\citenamefont{Barea and Iachello}(2009)}]{Barea09}
\bibinfo{author}{\bibfnamefont{J.}~\bibnamefont{Barea}} \bibnamefont{and}
  \bibinfo{author}{\bibfnamefont{F.}~\bibnamefont{Iachello}},
  \bibinfo{journal}{Phys. Rev. C} \textbf{\bibinfo{volume}{79}},
  \bibinfo{pages}{044301} (\bibinfo{year}{2009}).

\bibitem[{\citenamefont{Barea et~al.}(2012)\citenamefont{Barea, Kotila, and
  Iachello}}]{Barea12}
\bibinfo{author}{\bibfnamefont{J.}~\bibnamefont{Barea}},
  \bibinfo{author}{\bibfnamefont{J.}~\bibnamefont{Kotila}}, \bibnamefont{and}
  \bibinfo{author}{\bibfnamefont{F.}~\bibnamefont{Iachello}},
  \bibinfo{journal}{Phys. Rev. Lett.} \textbf{\bibinfo{volume}{109}},
  \bibinfo{pages}{042501} (\bibinfo{year}{2012}).

\bibitem[{\citenamefont{Barea et~al.}(2013)\citenamefont{Barea, Kotila, and
  Iachello}}]{Barea13}
\bibinfo{author}{\bibfnamefont{J.}~\bibnamefont{Barea}},
  \bibinfo{author}{\bibfnamefont{J.}~\bibnamefont{Kotila}}, \bibnamefont{and}
  \bibinfo{author}{\bibfnamefont{F.}~\bibnamefont{Iachello}},
  \bibinfo{journal}{Phys. Rev. C} \textbf{\bibinfo{volume}{87}},
  \bibinfo{pages}{014315} (\bibinfo{year}{2013}).

\bibitem[{\citenamefont{\ifmmode~\check{S}\else \v{S}\fi{}imkovic
  et~al.}(2008)\citenamefont{\ifmmode~\check{S}\else \v{S}\fi{}imkovic,
  Faessler, Rodin, Vogel, and Engel}}]{Simkovic08}
\bibinfo{author}{\bibfnamefont{F.}~\bibnamefont{\ifmmode~\check{S}\else
  \v{S}\fi{}imkovic}},
  \bibinfo{author}{\bibfnamefont{A.}~\bibnamefont{Faessler}},
  \bibinfo{author}{\bibfnamefont{V.}~\bibnamefont{Rodin}},
  \bibinfo{author}{\bibfnamefont{P.}~\bibnamefont{Vogel}}, \bibnamefont{and}
  \bibinfo{author}{\bibfnamefont{J.}~\bibnamefont{Engel}},
  \bibinfo{journal}{Phys. Rev. C} \textbf{\bibinfo{volume}{77}},
  \bibinfo{pages}{045503} (\bibinfo{year}{2008}).

\bibitem[{\citenamefont{\ifmmode~\check{S}\else \v{S}\fi{}imkovic
  et~al.}(2009)\citenamefont{\ifmmode~\check{S}\else \v{S}\fi{}imkovic,
  Faessler, M\"uther, Rodin, and Stauf}}]{Simkovic09}
\bibinfo{author}{\bibfnamefont{F.}~\bibnamefont{\ifmmode~\check{S}\else
  \v{S}\fi{}imkovic}},
  \bibinfo{author}{\bibfnamefont{A.}~\bibnamefont{Faessler}},
  \bibinfo{author}{\bibfnamefont{H.}~\bibnamefont{M\"uther}},
  \bibinfo{author}{\bibfnamefont{V.}~\bibnamefont{Rodin}}, \bibnamefont{and}
  \bibinfo{author}{\bibfnamefont{M.}~\bibnamefont{Stauf}},
  \bibinfo{journal}{Phys. Rev. C} \textbf{\bibinfo{volume}{79}},
  \bibinfo{pages}{055501} (\bibinfo{year}{2009}).

\bibitem[{\citenamefont{Fang et~al.}(2011)\citenamefont{Fang, Faessler, Rodin,
  and \ifmmode~\check{S}\else \v{S}\fi{}imkovic}}]{Fang11}
\bibinfo{author}{\bibfnamefont{D.-L.} \bibnamefont{Fang}},
  \bibinfo{author}{\bibfnamefont{A.}~\bibnamefont{Faessler}},
  \bibinfo{author}{\bibfnamefont{V.}~\bibnamefont{Rodin}}, \bibnamefont{and}
  \bibinfo{author}{\bibfnamefont{F.}~\bibnamefont{\ifmmode~\check{S}\else
  \v{S}\fi{}imkovic}}, \bibinfo{journal}{Phys. Rev. C}
  \textbf{\bibinfo{volume}{83}}, \bibinfo{pages}{034320}
  (\bibinfo{year}{2011}).

\bibitem[{\citenamefont{Faessler et~al.}(2012)\citenamefont{Faessler, Rodin,
  and Simkovic}}]{Faessler12}
\bibinfo{author}{\bibfnamefont{A.}~\bibnamefont{Faessler}},
  \bibinfo{author}{\bibfnamefont{V.}~\bibnamefont{Rodin}}, \bibnamefont{and}
  \bibinfo{author}{\bibfnamefont{F.}~\bibnamefont{Simkovic}},
  \bibinfo{journal}{J. Phys. G} \textbf{\bibinfo{volume}{39}},
  \bibinfo{pages}{124006} (\bibinfo{year}{2012}).

\bibitem[{\citenamefont{\ifmmode~\check{S}\else \v{S}\fi{}imkovic
  et~al.}(2018)\citenamefont{\ifmmode~\check{S}\else \v{S}\fi{}imkovic,
  Smetana, and Vogel}}]{Simkovic18}
\bibinfo{author}{\bibfnamefont{F.}~\bibnamefont{\ifmmode~\check{S}\else
  \v{S}\fi{}imkovic}},
  \bibinfo{author}{\bibfnamefont{A.}~\bibnamefont{Smetana}}, \bibnamefont{and}
  \bibinfo{author}{\bibfnamefont{P.}~\bibnamefont{Vogel}},
  \bibinfo{journal}{Phys. Rev. C} \textbf{\bibinfo{volume}{98}},
  \bibinfo{pages}{064325} (\bibinfo{year}{2018}).

\bibitem[{\citenamefont{Caurier et~al.}(2008)\citenamefont{Caurier, Nowacki,
  and Poves}}]{Caurier08}
\bibinfo{author}{\bibfnamefont{E.}~\bibnamefont{Caurier}},
  \bibinfo{author}{\bibfnamefont{F.}~\bibnamefont{Nowacki}}, \bibnamefont{and}
  \bibinfo{author}{\bibfnamefont{A.}~\bibnamefont{Poves}},
  \bibinfo{journal}{Eur. Phys. J. A} \textbf{\bibinfo{volume}{36}},
  \bibinfo{pages}{195} (\bibinfo{year}{2008}).

\bibitem[{\citenamefont{Men\'endez
  et~al.}(2009{\natexlab{a}})\citenamefont{Men\'endez, Poves, Caurier, and
  Nowacki}}]{Menendez09a}
\bibinfo{author}{\bibfnamefont{J.}~\bibnamefont{Men\'endez}},
  \bibinfo{author}{\bibfnamefont{A.}~\bibnamefont{Poves}},
  \bibinfo{author}{\bibfnamefont{E.}~\bibnamefont{Caurier}}, \bibnamefont{and}
  \bibinfo{author}{\bibfnamefont{F.}~\bibnamefont{Nowacki}},
  \bibinfo{journal}{Phys. Rev. C} \textbf{\bibinfo{volume}{80}},
  \bibinfo{pages}{048501} (\bibinfo{year}{2009}{\natexlab{a}}).

\bibitem[{\citenamefont{Men\'endez
  et~al.}(2009{\natexlab{b}})\citenamefont{Men\'endez, Poves, Caurier, and
  Nowacki}}]{Menendez09b}
\bibinfo{author}{\bibfnamefont{J.}~\bibnamefont{Men\'endez}},
  \bibinfo{author}{\bibfnamefont{A.}~\bibnamefont{Poves}},
  \bibinfo{author}{\bibfnamefont{E.}~\bibnamefont{Caurier}}, \bibnamefont{and}
  \bibinfo{author}{\bibfnamefont{F.}~\bibnamefont{Nowacki}},
  \bibinfo{journal}{Nucl. Phys. A} \textbf{\bibinfo{volume}{818}},
  \bibinfo{pages}{139} (\bibinfo{year}{2009}{\natexlab{b}}).

\bibitem[{\citenamefont{Caurier et~al.}(2012)\citenamefont{Caurier, Nowacki,
  and Poves}}]{Caurier12}
\bibinfo{author}{\bibfnamefont{E.}~\bibnamefont{Caurier}},
  \bibinfo{author}{\bibfnamefont{F.}~\bibnamefont{Nowacki}}, \bibnamefont{and}
  \bibinfo{author}{\bibfnamefont{A.}~\bibnamefont{Poves}},
  \bibinfo{journal}{Phys. Lett. B} \textbf{\bibinfo{volume}{711}},
  \bibinfo{pages}{62} (\bibinfo{year}{2012}).

\bibitem[{\citenamefont{Horoi et~al.}(2007)\citenamefont{Horoi, Stoica, and
  Brown}}]{Horoi07}
\bibinfo{author}{\bibfnamefont{M.}~\bibnamefont{Horoi}},
  \bibinfo{author}{\bibfnamefont{S.}~\bibnamefont{Stoica}}, \bibnamefont{and}
  \bibinfo{author}{\bibfnamefont{B.~A.} \bibnamefont{Brown}},
  \bibinfo{journal}{Phys. Rev. C} \textbf{\bibinfo{volume}{75}},
  \bibinfo{pages}{034303} (\bibinfo{year}{2007}).

\bibitem[{\citenamefont{Horoi}(2013{\natexlab{a}})}]{Horoi13a}
\bibinfo{author}{\bibfnamefont{M.}~\bibnamefont{Horoi}},
  \bibinfo{journal}{Phys. Rev. C} \textbf{\bibinfo{volume}{87}},
  \bibinfo{pages}{014320} (\bibinfo{year}{2013}{\natexlab{a}}).

\bibitem[{\citenamefont{Horoi and Brown}(2013)}]{Horoi13b}
\bibinfo{author}{\bibfnamefont{M.}~\bibnamefont{Horoi}} \bibnamefont{and}
  \bibinfo{author}{\bibfnamefont{B.~A.} \bibnamefont{Brown}},
  \bibinfo{journal}{Phys. Rev. Lett.} \textbf{\bibinfo{volume}{110}},
  \bibinfo{pages}{222502} (\bibinfo{year}{2013}).

\bibitem[{\citenamefont{Neacsu and Horoi}(2015)}]{Neacsu15}
\bibinfo{author}{\bibfnamefont{A.}~\bibnamefont{Neacsu}} \bibnamefont{and}
  \bibinfo{author}{\bibfnamefont{M.}~\bibnamefont{Horoi}},
  \bibinfo{journal}{Phys. Rev. C} \textbf{\bibinfo{volume}{91}},
  \bibinfo{pages}{024309} (\bibinfo{year}{2015}).

\bibitem[{\citenamefont{Brown et~al.}(2015)\citenamefont{Brown, Fang, and
  Horoi}}]{Brown15}
\bibinfo{author}{\bibfnamefont{B.~A.} \bibnamefont{Brown}},
  \bibinfo{author}{\bibfnamefont{D.~L.} \bibnamefont{Fang}}, \bibnamefont{and}
  \bibinfo{author}{\bibfnamefont{M.}~\bibnamefont{Horoi}},
  \bibinfo{journal}{Phys. Rev. C} \textbf{\bibinfo{volume}{92}},
  \bibinfo{pages}{041301} (\bibinfo{year}{2015}).

\bibitem[{\citenamefont{Iwata et~al.}(2016)\citenamefont{Iwata, Shimizu,
  Otsuka, Utsuno, Men\'endez, Honma, and Abe}}]{Iwata16}
\bibinfo{author}{\bibfnamefont{Y.}~\bibnamefont{Iwata}},
  \bibinfo{author}{\bibfnamefont{N.}~\bibnamefont{Shimizu}},
  \bibinfo{author}{\bibfnamefont{T.}~\bibnamefont{Otsuka}},
  \bibinfo{author}{\bibfnamefont{Y.}~\bibnamefont{Utsuno}},
  \bibinfo{author}{\bibfnamefont{J.}~\bibnamefont{Men\'endez}},
  \bibinfo{author}{\bibfnamefont{M.}~\bibnamefont{Honma}}, \bibnamefont{and}
  \bibinfo{author}{\bibfnamefont{T.}~\bibnamefont{Abe}},
  \bibinfo{journal}{Phys. Rev. Lett.} \textbf{\bibinfo{volume}{116}},
  \bibinfo{pages}{112502} (\bibinfo{year}{2016}).

\bibitem[{\citenamefont{Suhonen}(2017{\natexlab{a}})}]{Suhonen17b}
\bibinfo{author}{\bibfnamefont{J.}~\bibnamefont{Suhonen}},
  \bibinfo{journal}{Front. Phys.} \textbf{\bibinfo{volume}{5:55}}
  (\bibinfo{year}{2017}{\natexlab{a}}).

\bibitem[{\citenamefont{Park et~al.}(1993)\citenamefont{Park, Min, and
  Rho}}]{Park93}
\bibinfo{author}{\bibfnamefont{T.~S.} \bibnamefont{Park}},
  \bibinfo{author}{\bibfnamefont{D.~P.} \bibnamefont{Min}}, \bibnamefont{and}
  \bibinfo{author}{\bibfnamefont{M.}~\bibnamefont{Rho}},
  \bibinfo{journal}{Phys. Rep.} \textbf{\bibinfo{volume}{233}},
  \bibinfo{pages}{341} (\bibinfo{year}{1993}).

\bibitem[{\citenamefont{Pastore et~al.}(2009)\citenamefont{Pastore, Girlanda,
  Schiavilla, Viviani, and Wiringa}}]{Pastore09}
\bibinfo{author}{\bibfnamefont{S.}~\bibnamefont{Pastore}},
  \bibinfo{author}{\bibfnamefont{L.}~\bibnamefont{Girlanda}},
  \bibinfo{author}{\bibfnamefont{R.}~\bibnamefont{Schiavilla}},
  \bibinfo{author}{\bibfnamefont{M.}~\bibnamefont{Viviani}}, \bibnamefont{and}
  \bibinfo{author}{\bibfnamefont{R.~B.} \bibnamefont{Wiringa}},
  \bibinfo{journal}{Phys. Rev. C} \textbf{\bibinfo{volume}{80}},
  \bibinfo{pages}{034004} (\bibinfo{year}{2009}).

\bibitem[{\citenamefont{Piarulli et~al.}(2013)\citenamefont{Piarulli, Girlanda,
  Marcucci, Pastore, Schiavilla, and Viviani}}]{Piarulli13}
\bibinfo{author}{\bibfnamefont{M.}~\bibnamefont{Piarulli}},
  \bibinfo{author}{\bibfnamefont{L.}~\bibnamefont{Girlanda}},
  \bibinfo{author}{\bibfnamefont{L.~E.} \bibnamefont{Marcucci}},
  \bibinfo{author}{\bibfnamefont{S.}~\bibnamefont{Pastore}},
  \bibinfo{author}{\bibfnamefont{R.}~\bibnamefont{Schiavilla}},
  \bibnamefont{and} \bibinfo{author}{\bibfnamefont{M.}~\bibnamefont{Viviani}},
  \bibinfo{journal}{Phys. Rev. C} \textbf{\bibinfo{volume}{87}},
  \bibinfo{pages}{014006} (\bibinfo{year}{2013}).

\bibitem[{\citenamefont{Baroni et~al.}(2016{\natexlab{a}})\citenamefont{Baroni,
  Girlanda, Pastore, Schiavilla, and Viviani}}]{Baroni16b}
\bibinfo{author}{\bibfnamefont{A.}~\bibnamefont{Baroni}},
  \bibinfo{author}{\bibfnamefont{L.}~\bibnamefont{Girlanda}},
  \bibinfo{author}{\bibfnamefont{S.}~\bibnamefont{Pastore}},
  \bibinfo{author}{\bibfnamefont{R.}~\bibnamefont{Schiavilla}},
  \bibnamefont{and} \bibinfo{author}{\bibfnamefont{M.}~\bibnamefont{Viviani}},
  \bibinfo{journal}{Phys. Rev. C} \textbf{\bibinfo{volume}{93}},
  \bibinfo{pages}{015501} (\bibinfo{year}{2016}{\natexlab{a}}).

\bibitem[{\citenamefont{Tanabashi et~al.}(2018)\citenamefont{Tanabashi,
  Hagiwara, Hikasa, Nakamura, Sumino, Takahashi, Tanaka, Agashe, Aielli, Amsler
  et~al.}}]{PDB18}
\bibinfo{author}{\bibfnamefont{M.}~\bibnamefont{Tanabashi}},
  \bibinfo{author}{\bibfnamefont{K.}~\bibnamefont{Hagiwara}},
  \bibinfo{author}{\bibfnamefont{K.}~\bibnamefont{Hikasa}},
  \bibinfo{author}{\bibfnamefont{K.}~\bibnamefont{Nakamura}},
  \bibinfo{author}{\bibfnamefont{Y.}~\bibnamefont{Sumino}},
  \bibinfo{author}{\bibfnamefont{F.}~\bibnamefont{Takahashi}},
  \bibinfo{author}{\bibfnamefont{J.}~\bibnamefont{Tanaka}},
  \bibinfo{author}{\bibfnamefont{K.}~\bibnamefont{Agashe}},
  \bibinfo{author}{\bibfnamefont{G.}~\bibnamefont{Aielli}},
  \bibinfo{author}{\bibfnamefont{C.}~\bibnamefont{Amsler}},
  \bibnamefont{et~al.} (\bibinfo{collaboration}{Particle Data Group}),
  \bibinfo{journal}{Phys. Rev. D} \textbf{\bibinfo{volume}{98}},
  \bibinfo{pages}{030001} (\bibinfo{year}{2018}).

\bibitem[{\citenamefont{Towner}(1987)}]{Towner87}
\bibinfo{author}{\bibfnamefont{I.~S.} \bibnamefont{Towner}},
  \bibinfo{journal}{Phys. Rep.} \textbf{\bibinfo{volume}{155}},
  \bibinfo{pages}{263} (\bibinfo{year}{1987}).

\bibitem[{\citenamefont{Towner and Khanna}(1983)}]{Towner83}
\bibinfo{author}{\bibfnamefont{I.~S.} \bibnamefont{Towner}} \bibnamefont{and}
  \bibinfo{author}{\bibfnamefont{K.~F.~C.} \bibnamefont{Khanna}},
  \bibinfo{journal}{Nucl. Phys. A} \textbf{\bibinfo{volume}{399}},
  \bibinfo{pages}{334} (\bibinfo{year}{1983}).

\bibitem[{\citenamefont{Suhonen}(2017{\natexlab{b}})}]{Suhonen17a}
\bibinfo{author}{\bibfnamefont{J.}~\bibnamefont{Suhonen}},
  \bibinfo{journal}{Phys. Rev. C} \textbf{\bibinfo{volume}{96}},
  \bibinfo{pages}{055501} (\bibinfo{year}{2017}{\natexlab{b}}).

\bibitem[{\citenamefont{Baroni et~al.}(2016{\natexlab{b}})\citenamefont{Baroni,
  Girlanda, Kievsky, Marcucci, Schiavilla, and Viviani}}]{Baroni16a}
\bibinfo{author}{\bibfnamefont{A.}~\bibnamefont{Baroni}},
  \bibinfo{author}{\bibfnamefont{L.}~\bibnamefont{Girlanda}},
  \bibinfo{author}{\bibfnamefont{A.}~\bibnamefont{Kievsky}},
  \bibinfo{author}{\bibfnamefont{L.~E.} \bibnamefont{Marcucci}},
  \bibinfo{author}{\bibfnamefont{R.}~\bibnamefont{Schiavilla}},
  \bibnamefont{and} \bibinfo{author}{\bibfnamefont{M.}~\bibnamefont{Viviani}},
  \bibinfo{journal}{Phys. Rev. C} \textbf{\bibinfo{volume}{94}},
  \bibinfo{pages}{024003} (\bibinfo{year}{2016}{\natexlab{b}}).

\bibitem[{\citenamefont{Kuo et~al.}(1981)\citenamefont{Kuo, Shurpin, Tam,
  Osnes, and Ellis}}]{Kuo81}
\bibinfo{author}{\bibfnamefont{T.~T.~S.} \bibnamefont{Kuo}},
  \bibinfo{author}{\bibfnamefont{J.}~\bibnamefont{Shurpin}},
  \bibinfo{author}{\bibfnamefont{K.~C.} \bibnamefont{Tam}},
  \bibinfo{author}{\bibfnamefont{E.}~\bibnamefont{Osnes}}, \bibnamefont{and}
  \bibinfo{author}{\bibfnamefont{P.~J.} \bibnamefont{Ellis}},
  \bibinfo{journal}{Ann. Phys. (NY)} \textbf{\bibinfo{volume}{132}},
  \bibinfo{pages}{237} (\bibinfo{year}{1981}).

\bibitem[{\citenamefont{Coraggio et~al.}(2017)\citenamefont{Coraggio,
  De~Angelis, Fukui, Gargano, and Itaco}}]{Coraggio17a}
\bibinfo{author}{\bibfnamefont{L.}~\bibnamefont{Coraggio}},
  \bibinfo{author}{\bibfnamefont{L.}~\bibnamefont{De~Angelis}},
  \bibinfo{author}{\bibfnamefont{T.}~\bibnamefont{Fukui}},
  \bibinfo{author}{\bibfnamefont{A.}~\bibnamefont{Gargano}}, \bibnamefont{and}
  \bibinfo{author}{\bibfnamefont{N.}~\bibnamefont{Itaco}},
  \bibinfo{journal}{Phys. Rev. C} \textbf{\bibinfo{volume}{95}},
  \bibinfo{pages}{064324} (\bibinfo{year}{2017}).

\bibitem[{\citenamefont{Machleidt}(2001)}]{Machleidt01b}
\bibinfo{author}{\bibfnamefont{R.}~\bibnamefont{Machleidt}},
  \bibinfo{journal}{Phys. Rev. C} \textbf{\bibinfo{volume}{63}},
  \bibinfo{pages}{024001} (\bibinfo{year}{2001}).

\bibitem[{\citenamefont{Bogner et~al.}(2001)\citenamefont{Bogner, Kuo, and
  Coraggio}}]{Bogner01}
\bibinfo{author}{\bibfnamefont{S.}~\bibnamefont{Bogner}},
  \bibinfo{author}{\bibfnamefont{T.~T.~S.} \bibnamefont{Kuo}},
  \bibnamefont{and} \bibinfo{author}{\bibfnamefont{L.}~\bibnamefont{Coraggio}},
  \bibinfo{journal}{Nucl. Phys. A} \textbf{\bibinfo{volume}{684}},
  \bibinfo{pages}{432c} (\bibinfo{year}{2001}).

\bibitem[{\citenamefont{Bogner et~al.}(2002)\citenamefont{Bogner, Kuo,
  Coraggio, Covello, and Itaco}}]{Bogner02}
\bibinfo{author}{\bibfnamefont{S.}~\bibnamefont{Bogner}},
  \bibinfo{author}{\bibfnamefont{T.~T.~S.} \bibnamefont{Kuo}},
  \bibinfo{author}{\bibfnamefont{L.}~\bibnamefont{Coraggio}},
  \bibinfo{author}{\bibfnamefont{A.}~\bibnamefont{Covello}}, \bibnamefont{and}
  \bibinfo{author}{\bibfnamefont{N.}~\bibnamefont{Itaco}},
  \bibinfo{journal}{Phys. Rev. C} \textbf{\bibinfo{volume}{65}},
  \bibinfo{pages}{051301(R)} (\bibinfo{year}{2002}).

\bibitem[{\citenamefont{Coraggio
  et~al.}(2015{\natexlab{a}})\citenamefont{Coraggio, Covello, Gargano, Itaco,
  and Kuo}}]{Coraggio15a}
\bibinfo{author}{\bibfnamefont{L.}~\bibnamefont{Coraggio}},
  \bibinfo{author}{\bibfnamefont{A.}~\bibnamefont{Covello}},
  \bibinfo{author}{\bibfnamefont{A.}~\bibnamefont{Gargano}},
  \bibinfo{author}{\bibfnamefont{N.}~\bibnamefont{Itaco}}, \bibnamefont{and}
  \bibinfo{author}{\bibfnamefont{T.~T.~S.} \bibnamefont{Kuo}},
  \bibinfo{journal}{Phys. Rev. C} \textbf{\bibinfo{volume}{91}},
  \bibinfo{pages}{041301} (\bibinfo{year}{2015}{\natexlab{a}}).

\bibitem[{\citenamefont{Coraggio
  et~al.}(2015{\natexlab{b}})\citenamefont{Coraggio, Gargano, and
  Itaco}}]{Coraggio15b}
\bibinfo{author}{\bibfnamefont{L.}~\bibnamefont{Coraggio}},
  \bibinfo{author}{\bibfnamefont{A.}~\bibnamefont{Gargano}}, \bibnamefont{and}
  \bibinfo{author}{\bibfnamefont{N.}~\bibnamefont{Itaco}},
  \bibinfo{journal}{JPS Conf. Proc.} \textbf{\bibinfo{volume}{6}},
  \bibinfo{pages}{020046} (\bibinfo{year}{2015}{\natexlab{b}}).

\bibitem[{\citenamefont{Coraggio et~al.}(2016)\citenamefont{Coraggio, Gargano,
  and Itaco}}]{Coraggio16a}
\bibinfo{author}{\bibfnamefont{L.}~\bibnamefont{Coraggio}},
  \bibinfo{author}{\bibfnamefont{A.}~\bibnamefont{Gargano}}, \bibnamefont{and}
  \bibinfo{author}{\bibfnamefont{N.}~\bibnamefont{Itaco}},
  \bibinfo{journal}{Phys. Rev. C} \textbf{\bibinfo{volume}{93}},
  \bibinfo{pages}{064328} (\bibinfo{year}{2016}).

\bibitem[{\citenamefont{Kuo and Osnes}(1990)}]{Kuo90}
\bibinfo{author}{\bibfnamefont{T.~T.~S.} \bibnamefont{Kuo}} \bibnamefont{and}
  \bibinfo{author}{\bibfnamefont{E.}~\bibnamefont{Osnes}},
  \emph{\bibinfo{title}{Lecture Notes in Physics}}, vol. \bibinfo{volume}{364}
  (\bibinfo{publisher}{Springer-Verlag, Berlin}, \bibinfo{year}{1990}).

\bibitem[{\citenamefont{Kuo et~al.}(1995)\citenamefont{Kuo, Krmpoti\'c, Suzuki,
  and Okamoto}}]{Kuo95}
\bibinfo{author}{\bibfnamefont{T.~T.~S.} \bibnamefont{Kuo}},
  \bibinfo{author}{\bibfnamefont{F.}~\bibnamefont{Krmpoti\'c}},
  \bibinfo{author}{\bibfnamefont{K.}~\bibnamefont{Suzuki}}, \bibnamefont{and}
  \bibinfo{author}{\bibfnamefont{R.}~\bibnamefont{Okamoto}},
  \bibinfo{journal}{Nucl. Phys. A} \textbf{\bibinfo{volume}{582}},
  \bibinfo{pages}{205} (\bibinfo{year}{1995}).

\bibitem[{\citenamefont{Kuo et~al.}(1971)\citenamefont{Kuo, Lee, and
  Ratcliff}}]{Kuo71}
\bibinfo{author}{\bibfnamefont{T.~T.~S.} \bibnamefont{Kuo}},
  \bibinfo{author}{\bibfnamefont{S.~Y.} \bibnamefont{Lee}}, \bibnamefont{and}
  \bibinfo{author}{\bibfnamefont{K.~F.} \bibnamefont{Ratcliff}},
  \bibinfo{journal}{Nucl. Phys. A} \textbf{\bibinfo{volume}{176}},
  \bibinfo{pages}{65} (\bibinfo{year}{1971}).

\bibitem[{\citenamefont{Coraggio et~al.}(2010)\citenamefont{Coraggio, Covello,
  Gargano, and Itaco}}]{Coraggio10a}
\bibinfo{author}{\bibfnamefont{L.}~\bibnamefont{Coraggio}},
  \bibinfo{author}{\bibfnamefont{A.}~\bibnamefont{Covello}},
  \bibinfo{author}{\bibfnamefont{A.}~\bibnamefont{Gargano}}, \bibnamefont{and}
  \bibinfo{author}{\bibfnamefont{N.}~\bibnamefont{Itaco}},
  \bibinfo{journal}{Phys. Rev. C} \textbf{\bibinfo{volume}{81}},
  \bibinfo{pages}{064303} (\bibinfo{year}{2010}).

\bibitem[{\citenamefont{Brandow}(1967)}]{Brandow67}
\bibinfo{author}{\bibfnamefont{B.~H.} \bibnamefont{Brandow}},
  \bibinfo{journal}{Rev. Mod. Phys.} \textbf{\bibinfo{volume}{39}},
  \bibinfo{pages}{771} (\bibinfo{year}{1967}).

\bibitem[{\citenamefont{Krenciglowa and Kuo}(1974)}]{Krenciglowa74}
\bibinfo{author}{\bibfnamefont{E.~M.} \bibnamefont{Krenciglowa}}
  \bibnamefont{and} \bibinfo{author}{\bibfnamefont{T.~T.~S.}
  \bibnamefont{Kuo}}, \bibinfo{journal}{Nucl. Phys. A}
  \textbf{\bibinfo{volume}{235}}, \bibinfo{pages}{171} (\bibinfo{year}{1974}).

\bibitem[{sup()}]{supplemental2018}
\bibinfo{note}{See Supplemental material at [URL will be inserted by publisher]
  for the list of two-body matrix elements of the shell-model hamiltonian
  $H_{\rm eff}$}.

\bibitem[{\citenamefont{Coraggio et~al.}(2018)\citenamefont{Coraggio,
  De~Angelis, Fukui, Gargano, and Itaco}}]{Coraggio18b}
\bibinfo{author}{\bibfnamefont{L.}~\bibnamefont{Coraggio}},
  \bibinfo{author}{\bibfnamefont{L.}~\bibnamefont{De~Angelis}},
  \bibinfo{author}{\bibfnamefont{T.}~\bibnamefont{Fukui}},
  \bibinfo{author}{\bibfnamefont{A.}~\bibnamefont{Gargano}}, \bibnamefont{and}
  \bibinfo{author}{\bibfnamefont{N.}~\bibnamefont{Itaco}}, \bibinfo{journal}{J.
  Phys. Conf. Ser.} \textbf{\bibinfo{volume}{1056}}, \bibinfo{pages}{012012}
  (\bibinfo{year}{2018}).

\bibitem[{\citenamefont{Baker and Gammel}(1970)}]{Baker70}
\bibinfo{author}{\bibfnamefont{G.~A.} \bibnamefont{Baker}} \bibnamefont{and}
  \bibinfo{author}{\bibfnamefont{J.~L.} \bibnamefont{Gammel}},
  \emph{\bibinfo{title}{The Pad{\'e} Approximant in Theoretical Physics}},
  vol.~\bibinfo{volume}{71} of \emph{\bibinfo{series}{Mathematics in Science
  and Engineering}} (\bibinfo{publisher}{Academic Press, New York},
  \bibinfo{year}{1970}).

\bibitem[{\citenamefont{Hoffmann et~al.}(1976)\citenamefont{Hoffmann, Starkand,
  and Kirson}}]{Hoffmann76}
\bibinfo{author}{\bibfnamefont{H.~M.} \bibnamefont{Hoffmann}},
  \bibinfo{author}{\bibfnamefont{Y.}~\bibnamefont{Starkand}}, \bibnamefont{and}
  \bibinfo{author}{\bibfnamefont{M.~W.} \bibnamefont{Kirson}},
  \bibinfo{journal}{Nucl. Phys. A} \textbf{\bibinfo{volume}{266}},
  \bibinfo{pages}{138} (\bibinfo{year}{1976}).

\bibitem[{\citenamefont{Goodman et~al.}(1980)\citenamefont{Goodman, Goulding,
  Greenfield, Rapaport, Bainum, Foster, Love, and Petrovich}}]{Goodman80}
\bibinfo{author}{\bibfnamefont{C.~D.} \bibnamefont{Goodman}},
  \bibinfo{author}{\bibfnamefont{C.~A.} \bibnamefont{Goulding}},
  \bibinfo{author}{\bibfnamefont{M.~B.} \bibnamefont{Greenfield}},
  \bibinfo{author}{\bibfnamefont{J.}~\bibnamefont{Rapaport}},
  \bibinfo{author}{\bibfnamefont{D.~E.} \bibnamefont{Bainum}},
  \bibinfo{author}{\bibfnamefont{C.~C.} \bibnamefont{Foster}},
  \bibinfo{author}{\bibfnamefont{W.~G.} \bibnamefont{Love}}, \bibnamefont{and}
  \bibinfo{author}{\bibfnamefont{F.}~\bibnamefont{Petrovich}},
  \bibinfo{journal}{Phys. Rev. Lett.} \textbf{\bibinfo{volume}{44}},
  \bibinfo{pages}{1755} (\bibinfo{year}{1980}).

\bibitem[{\citenamefont{Taddeucci et~al.}(1987)\citenamefont{Taddeucci,
  Goulding, Carey, Byrd, Goodman, Gaarde, Larsen, Horen, Rapaport, and
  Sugarbaker}}]{Taddeucci87}
\bibinfo{author}{\bibfnamefont{T.~N.} \bibnamefont{Taddeucci}},
  \bibinfo{author}{\bibfnamefont{C.~A.} \bibnamefont{Goulding}},
  \bibinfo{author}{\bibfnamefont{T.~A.} \bibnamefont{Carey}},
  \bibinfo{author}{\bibfnamefont{R.~C.} \bibnamefont{Byrd}},
  \bibinfo{author}{\bibfnamefont{C.~D.} \bibnamefont{Goodman}},
  \bibinfo{author}{\bibfnamefont{C.}~\bibnamefont{Gaarde}},
  \bibinfo{author}{\bibfnamefont{J.}~\bibnamefont{Larsen}},
  \bibinfo{author}{\bibfnamefont{D.}~\bibnamefont{Horen}},
  \bibinfo{author}{\bibfnamefont{J.}~\bibnamefont{Rapaport}}, \bibnamefont{and}
  \bibinfo{author}{\bibfnamefont{E.}~\bibnamefont{Sugarbaker}},
  \bibinfo{journal}{Nucl. Phys. A} \textbf{\bibinfo{volume}{469}},
  \bibinfo{pages}{125} (\bibinfo{year}{1987}).

\bibitem[{\citenamefont{Caurier et~al.}(2005)\citenamefont{Caurier,
  Mart\'{\i}nez-Pinedo, Nowacki, Poves, and Zuker}}]{Caurier05}
\bibinfo{author}{\bibfnamefont{E.}~\bibnamefont{Caurier}},
  \bibinfo{author}{\bibfnamefont{G.}~\bibnamefont{Mart\'{\i}nez-Pinedo}},
  \bibinfo{author}{\bibfnamefont{F.}~\bibnamefont{Nowacki}},
  \bibinfo{author}{\bibfnamefont{A.}~\bibnamefont{Poves}}, \bibnamefont{and}
  \bibinfo{author}{\bibfnamefont{A.~P.} \bibnamefont{Zuker}},
  \bibinfo{journal}{Rev. Mod. Phys.} \textbf{\bibinfo{volume}{77}},
  \bibinfo{pages}{427} (\bibinfo{year}{2005}).

\bibitem[{\citenamefont{Haxton and Stephenson~Jr.}(1984)}]{Haxton84}
\bibinfo{author}{\bibfnamefont{W.~C.} \bibnamefont{Haxton}} \bibnamefont{and}
  \bibinfo{author}{\bibfnamefont{G.~J.} \bibnamefont{Stephenson~Jr.}},
  \bibinfo{journal}{Prog. Part. Nucl. Phys.} \textbf{\bibinfo{volume}{12}},
  \bibinfo{pages}{409} (\bibinfo{year}{1984}).

\bibitem[{\citenamefont{Sen'kov and Horoi}(2013)}]{Senkov13}
\bibinfo{author}{\bibfnamefont{R.~A.} \bibnamefont{Sen'kov}} \bibnamefont{and}
  \bibinfo{author}{\bibfnamefont{M.}~\bibnamefont{Horoi}},
  \bibinfo{journal}{Phys. Rev. C} \textbf{\bibinfo{volume}{88}},
  \bibinfo{pages}{064312} (\bibinfo{year}{2013}).

\bibitem[{\citenamefont{Horoi}(2013{\natexlab{b}})}]{Horoi13c}
\bibinfo{author}{\bibfnamefont{M.}~\bibnamefont{Horoi}}, \bibinfo{journal}{J.
  Phys. Conf. Ser.} \textbf{\bibinfo{volume}{413}}, \bibinfo{pages}{012020}
  (\bibinfo{year}{2013}{\natexlab{b}}).

\bibitem[{\citenamefont{Barabash}(2015)}]{Barabash15}
\bibinfo{author}{\bibfnamefont{A.~S.} \bibnamefont{Barabash}},
  \bibinfo{journal}{Nucl. Phys. A} \textbf{\bibinfo{volume}{935}},
  \bibinfo{pages}{52} (\bibinfo{year}{2015}).

\bibitem[{ens()}]{ensdf}
\bibinfo{note}{Data extracted using the NNDC On-line Data Service from the
  ENSDF database, file revised as of November 6, 2018.},
  \urlprefix\url{https://www.nndc.bnl.gov/ensdf}.

\bibitem[{xun()}]{xundl}
\bibinfo{note}{Data extracted using the NNDC On-line Data Service from the
  XUNDL database, file revised as of November 6, 2018.},
  \urlprefix\url{https://www.nndc.bnl.gov/ensdf/ensdf/xundl.jsp}.

\bibitem[{\citenamefont{Vanhoy et~al.}(1992)\citenamefont{Vanhoy, McEllistrem,
  Hicks, Gatenby, Baum, Johnson, Moln\'ar, and Yates}}]{Vanhoy92}
\bibinfo{author}{\bibfnamefont{J.~R.} \bibnamefont{Vanhoy}},
  \bibinfo{author}{\bibfnamefont{M.~T.} \bibnamefont{McEllistrem}},
  \bibinfo{author}{\bibfnamefont{S.~F.} \bibnamefont{Hicks}},
  \bibinfo{author}{\bibfnamefont{R.~A.} \bibnamefont{Gatenby}},
  \bibinfo{author}{\bibfnamefont{E.~M.} \bibnamefont{Baum}},
  \bibinfo{author}{\bibfnamefont{E.~L.} \bibnamefont{Johnson}},
  \bibinfo{author}{\bibfnamefont{G.}~\bibnamefont{Moln\'ar}}, \bibnamefont{and}
  \bibinfo{author}{\bibfnamefont{S.~W.} \bibnamefont{Yates}},
  \bibinfo{journal}{Phys. Rev. C} \textbf{\bibinfo{volume}{45}},
  \bibinfo{pages}{1628} (\bibinfo{year}{1992}).

\bibitem[{\citenamefont{Yako et~al.}(2009)\citenamefont{Yako, Sasano, Miki,
  Sakai, Dozono, Frekers, Greenfield, Hatanaka, Ihara, Kato et~al.}}]{Yako09}
\bibinfo{author}{\bibfnamefont{K.}~\bibnamefont{Yako}},
  \bibinfo{author}{\bibfnamefont{M.}~\bibnamefont{Sasano}},
  \bibinfo{author}{\bibfnamefont{K.}~\bibnamefont{Miki}},
  \bibinfo{author}{\bibfnamefont{H.}~\bibnamefont{Sakai}},
  \bibinfo{author}{\bibfnamefont{M.}~\bibnamefont{Dozono}},
  \bibinfo{author}{\bibfnamefont{D.}~\bibnamefont{Frekers}},
  \bibinfo{author}{\bibfnamefont{M.~B.} \bibnamefont{Greenfield}},
  \bibinfo{author}{\bibfnamefont{K.}~\bibnamefont{Hatanaka}},
  \bibinfo{author}{\bibfnamefont{E.}~\bibnamefont{Ihara}},
  \bibinfo{author}{\bibfnamefont{M.}~\bibnamefont{Kato}}, \bibnamefont{et~al.},
  \bibinfo{journal}{Phys. Rev. Lett.} \textbf{\bibinfo{volume}{103}},
  \bibinfo{pages}{012503} (\bibinfo{year}{2009}).

\bibitem[{\citenamefont{Mukhopadhyay et~al.}(2017)\citenamefont{Mukhopadhyay,
  Crider, Brown, Ashley, Chakraborty, Kumar, McEllistrem, Peters,
  Prados-Est\'evez, and Yates}}]{Mukhopadhyay17}
\bibinfo{author}{\bibfnamefont{S.}~\bibnamefont{Mukhopadhyay}},
  \bibinfo{author}{\bibfnamefont{B.~P.} \bibnamefont{Crider}},
  \bibinfo{author}{\bibfnamefont{B.~A.} \bibnamefont{Brown}},
  \bibinfo{author}{\bibfnamefont{S.~F.} \bibnamefont{Ashley}},
  \bibinfo{author}{\bibfnamefont{A.}~\bibnamefont{Chakraborty}},
  \bibinfo{author}{\bibfnamefont{A.}~\bibnamefont{Kumar}},
  \bibinfo{author}{\bibfnamefont{M.~T.} \bibnamefont{McEllistrem}},
  \bibinfo{author}{\bibfnamefont{E.~E.} \bibnamefont{Peters}},
  \bibinfo{author}{\bibfnamefont{F.~M.} \bibnamefont{Prados-Est\'evez}},
  \bibnamefont{and} \bibinfo{author}{\bibfnamefont{S.~W.} \bibnamefont{Yates}},
  \bibinfo{journal}{Phys. Rev. C} \textbf{\bibinfo{volume}{95}},
  \bibinfo{pages}{014327} (\bibinfo{year}{2017}).

\bibitem[{\citenamefont{G\"urdal et~al.}(2013)\citenamefont{G\"urdal,
  Stefanova, Boutachkov, Torres, Kumbartzki, Benczer-Koller, Sharon, Zamick,
  Robinson, Ahn et~al.}}]{Gurdal13}
\bibinfo{author}{\bibfnamefont{G.}~\bibnamefont{G\"urdal}},
  \bibinfo{author}{\bibfnamefont{E.~A.} \bibnamefont{Stefanova}},
  \bibinfo{author}{\bibfnamefont{P.}~\bibnamefont{Boutachkov}},
  \bibinfo{author}{\bibfnamefont{D.~A.} \bibnamefont{Torres}},
  \bibinfo{author}{\bibfnamefont{G.~J.} \bibnamefont{Kumbartzki}},
  \bibinfo{author}{\bibfnamefont{N.}~\bibnamefont{Benczer-Koller}},
  \bibinfo{author}{\bibfnamefont{Y.~Y.} \bibnamefont{Sharon}},
  \bibinfo{author}{\bibfnamefont{L.}~\bibnamefont{Zamick}},
  \bibinfo{author}{\bibfnamefont{S.~J.~Q.} \bibnamefont{Robinson}},
  \bibinfo{author}{\bibfnamefont{T.}~\bibnamefont{Ahn}}, \bibnamefont{et~al.},
  \bibinfo{journal}{Phys. Rev. C} \textbf{\bibinfo{volume}{88}},
  \bibinfo{pages}{014301} (\bibinfo{year}{2013}).

\bibitem[{\citenamefont{Speidel et~al.}(1998)\citenamefont{Speidel,
  Benczer-Koller, Kumbartzki, Barton, Gelberg, Holden, Jakob, Matt, Mayer,
  Satteson et~al.}}]{Speidel98}
\bibinfo{author}{\bibfnamefont{K.-H.} \bibnamefont{Speidel}},
  \bibinfo{author}{\bibfnamefont{N.}~\bibnamefont{Benczer-Koller}},
  \bibinfo{author}{\bibfnamefont{G.}~\bibnamefont{Kumbartzki}},
  \bibinfo{author}{\bibfnamefont{C.}~\bibnamefont{Barton}},
  \bibinfo{author}{\bibfnamefont{A.}~\bibnamefont{Gelberg}},
  \bibinfo{author}{\bibfnamefont{J.}~\bibnamefont{Holden}},
  \bibinfo{author}{\bibfnamefont{G.}~\bibnamefont{Jakob}},
  \bibinfo{author}{\bibfnamefont{N.}~\bibnamefont{Matt}},
  \bibinfo{author}{\bibfnamefont{R.~H.} \bibnamefont{Mayer}},
  \bibinfo{author}{\bibfnamefont{M.}~\bibnamefont{Satteson}},
  \bibnamefont{et~al.}, \bibinfo{journal}{Phys. Rev. C}
  \textbf{\bibinfo{volume}{57}}, \bibinfo{pages}{2181} (\bibinfo{year}{1998}).

\bibitem[{\citenamefont{Thies et~al.}(2012)\citenamefont{Thies, Frekers,
  Adachi, Dozono, Ejiri, Fujita, Fujita, Fujiwara, Grewe, Hatanaka
  et~al.}}]{Thies12a}
\bibinfo{author}{\bibfnamefont{J.~H.} \bibnamefont{Thies}},
  \bibinfo{author}{\bibfnamefont{D.}~\bibnamefont{Frekers}},
  \bibinfo{author}{\bibfnamefont{T.}~\bibnamefont{Adachi}},
  \bibinfo{author}{\bibfnamefont{M.}~\bibnamefont{Dozono}},
  \bibinfo{author}{\bibfnamefont{H.}~\bibnamefont{Ejiri}},
  \bibinfo{author}{\bibfnamefont{H.}~\bibnamefont{Fujita}},
  \bibinfo{author}{\bibfnamefont{Y.}~\bibnamefont{Fujita}},
  \bibinfo{author}{\bibfnamefont{M.}~\bibnamefont{Fujiwara}},
  \bibinfo{author}{\bibfnamefont{E.-W.} \bibnamefont{Grewe}},
  \bibinfo{author}{\bibfnamefont{K.}~\bibnamefont{Hatanaka}},
  \bibnamefont{et~al.}, \bibinfo{journal}{Phys. Rev. C}
  \textbf{\bibinfo{volume}{86}}, \bibinfo{pages}{014304}
  (\bibinfo{year}{2012}).

\bibitem[{\citenamefont{Frekers et~al.}(2016)\citenamefont{Frekers, Alanssari,
  Adachi, Cleveland, Dozono, Ejiri, Elliott, Fujita, Fujita, Fujiwara
  et~al.}}]{Frekers16}
\bibinfo{author}{\bibfnamefont{D.}~\bibnamefont{Frekers}},
  \bibinfo{author}{\bibfnamefont{M.}~\bibnamefont{Alanssari}},
  \bibinfo{author}{\bibfnamefont{T.}~\bibnamefont{Adachi}},
  \bibinfo{author}{\bibfnamefont{B.~T.} \bibnamefont{Cleveland}},
  \bibinfo{author}{\bibfnamefont{M.}~\bibnamefont{Dozono}},
  \bibinfo{author}{\bibfnamefont{H.}~\bibnamefont{Ejiri}},
  \bibinfo{author}{\bibfnamefont{S.~R.} \bibnamefont{Elliott}},
  \bibinfo{author}{\bibfnamefont{H.}~\bibnamefont{Fujita}},
  \bibinfo{author}{\bibfnamefont{Y.}~\bibnamefont{Fujita}},
  \bibinfo{author}{\bibfnamefont{M.}~\bibnamefont{Fujiwara}},
  \bibnamefont{et~al.}, \bibinfo{journal}{Phys. Rev. C}
  \textbf{\bibinfo{volume}{94}}, \bibinfo{pages}{014614}
  (\bibinfo{year}{2016}).

\bibitem[{\citenamefont{Hicks et~al.}(2008)\citenamefont{Hicks, Vanhoy, and
  Yates}}]{Hicks08}
\bibinfo{author}{\bibfnamefont{S.~F.} \bibnamefont{Hicks}},
  \bibinfo{author}{\bibfnamefont{J.~R.} \bibnamefont{Vanhoy}},
  \bibnamefont{and} \bibinfo{author}{\bibfnamefont{S.~W.} \bibnamefont{Yates}},
  \bibinfo{journal}{Phys. Rev. C} \textbf{\bibinfo{volume}{78}},
  \bibinfo{pages}{054320} (\bibinfo{year}{2008}).

\bibitem[{\citenamefont{Puppe et~al.}(2012)\citenamefont{Puppe, Lennarz,
  Adachi, Akimune, Ejiri, Frekers, Fujita, Fujita, Fujiwara,
  Ganio\ifmmode~\breve{g}\else \u{g}\fi{}lu et~al.}}]{Puppe12}
\bibinfo{author}{\bibfnamefont{P.}~\bibnamefont{Puppe}},
  \bibinfo{author}{\bibfnamefont{A.}~\bibnamefont{Lennarz}},
  \bibinfo{author}{\bibfnamefont{T.}~\bibnamefont{Adachi}},
  \bibinfo{author}{\bibfnamefont{H.}~\bibnamefont{Akimune}},
  \bibinfo{author}{\bibfnamefont{H.}~\bibnamefont{Ejiri}},
  \bibinfo{author}{\bibfnamefont{D.}~\bibnamefont{Frekers}},
  \bibinfo{author}{\bibfnamefont{H.}~\bibnamefont{Fujita}},
  \bibinfo{author}{\bibfnamefont{Y.}~\bibnamefont{Fujita}},
  \bibinfo{author}{\bibfnamefont{M.}~\bibnamefont{Fujiwara}},
  \bibinfo{author}{\bibfnamefont{E.}~\bibnamefont{Ganio\ifmmode~\breve{g}\else
  \u{g}\fi{}lu}}, \bibnamefont{et~al.}, \bibinfo{journal}{Phys. Rev. C}
  \textbf{\bibinfo{volume}{86}}, \bibinfo{pages}{044603}
  (\bibinfo{year}{2012}).

\bibitem[{\citenamefont{Mukhopadhyay et~al.}(2008)\citenamefont{Mukhopadhyay,
  Scheck, Crider, Choudry, Elhami, Peters, McEllistrem, Orce, and
  Yates}}]{Mukhopadhyay08}
\bibinfo{author}{\bibfnamefont{S.}~\bibnamefont{Mukhopadhyay}},
  \bibinfo{author}{\bibfnamefont{M.}~\bibnamefont{Scheck}},
  \bibinfo{author}{\bibfnamefont{B.}~\bibnamefont{Crider}},
  \bibinfo{author}{\bibfnamefont{S.~N.} \bibnamefont{Choudry}},
  \bibinfo{author}{\bibfnamefont{E.}~\bibnamefont{Elhami}},
  \bibinfo{author}{\bibfnamefont{E.}~\bibnamefont{Peters}},
  \bibinfo{author}{\bibfnamefont{M.~T.} \bibnamefont{McEllistrem}},
  \bibinfo{author}{\bibfnamefont{J.~N.} \bibnamefont{Orce}}, \bibnamefont{and}
  \bibinfo{author}{\bibfnamefont{S.~W.} \bibnamefont{Yates}},
  \bibinfo{journal}{Phys. Rev. C} \textbf{\bibinfo{volume}{78}},
  \bibinfo{pages}{034317} (\bibinfo{year}{2008}).

\bibitem[{\citenamefont{Frekers et~al.}(2013)\citenamefont{Frekers, Puppe,
  Thies, and Ejiri}}]{Frekers13}
\bibinfo{author}{\bibfnamefont{D.}~\bibnamefont{Frekers}},
  \bibinfo{author}{\bibfnamefont{P.}~\bibnamefont{Puppe}},
  \bibinfo{author}{\bibfnamefont{J.~H.} \bibnamefont{Thies}}, \bibnamefont{and}
  \bibinfo{author}{\bibfnamefont{H.}~\bibnamefont{Ejiri}},
  \bibinfo{journal}{Nucl. Phys. A} \textbf{\bibinfo{volume}{916}},
  \bibinfo{pages}{219} (\bibinfo{year}{2013}).

\bibitem[{\citenamefont{Johnstone and Towner}(1998)}]{Johnstone98}
\bibinfo{author}{\bibfnamefont{I.~P.} \bibnamefont{Johnstone}}
  \bibnamefont{and} \bibinfo{author}{\bibfnamefont{I.~S.}
  \bibnamefont{Towner}}, \bibinfo{journal}{Eur. Phys. J. A}
  \textbf{\bibinfo{volume}{3}}, \bibinfo{pages}{237} (\bibinfo{year}{1998}).

\bibitem[{\citenamefont{Brown et~al.}(2005)\citenamefont{Brown, Stone, Stone,
  Towner, and Hjorth-Jensen}}]{Brown05}
\bibinfo{author}{\bibfnamefont{B.~A.} \bibnamefont{Brown}},
  \bibinfo{author}{\bibfnamefont{N.~J.} \bibnamefont{Stone}},
  \bibinfo{author}{\bibfnamefont{J.~R.} \bibnamefont{Stone}},
  \bibinfo{author}{\bibfnamefont{I.~S.} \bibnamefont{Towner}},
  \bibnamefont{and}
  \bibinfo{author}{\bibfnamefont{M.}~\bibnamefont{Hjorth-Jensen}},
  \bibinfo{journal}{Phys. Rev. C} \textbf{\bibinfo{volume}{71}},
  \bibinfo{pages}{044317} (\bibinfo{year}{2005}).

\bibitem[{\citenamefont{Fukui et~al.}(2018)\citenamefont{Fukui, De~Angelis, Ma,
  Coraggio, Gargano, Itaco, and Xu}}]{Coraggio18a}
\bibinfo{author}{\bibfnamefont{T.}~\bibnamefont{Fukui}},
  \bibinfo{author}{\bibfnamefont{L.}~\bibnamefont{De~Angelis}},
  \bibinfo{author}{\bibfnamefont{Y.~Z.} \bibnamefont{Ma}},
  \bibinfo{author}{\bibfnamefont{L.}~\bibnamefont{Coraggio}},
  \bibinfo{author}{\bibfnamefont{A.}~\bibnamefont{Gargano}},
  \bibinfo{author}{\bibfnamefont{N.}~\bibnamefont{Itaco}}, \bibnamefont{and}
  \bibinfo{author}{\bibfnamefont{F.~R.} \bibnamefont{Xu}},
  \bibinfo{journal}{Phys. Rev. C} \textbf{\bibinfo{volume}{98}},
  \bibinfo{pages}{044305} (\bibinfo{year}{2018}).

\bibitem[{\citenamefont{Men\'endez et~al.}(2011)\citenamefont{Men\'endez,
  Gazit, and Schwenk}}]{Menendez11}
\bibinfo{author}{\bibfnamefont{J.}~\bibnamefont{Men\'endez}},
  \bibinfo{author}{\bibfnamefont{D.}~\bibnamefont{Gazit}}, \bibnamefont{and}
  \bibinfo{author}{\bibfnamefont{A.}~\bibnamefont{Schwenk}},
  \bibinfo{journal}{Phys. Rev. Lett.} \textbf{\bibinfo{volume}{107}},
  \bibinfo{pages}{062501} (\bibinfo{year}{2011}).

\bibitem[{\citenamefont{Wang et~al.}(2018)\citenamefont{Wang, Engel, and
  Yao}}]{Wang18}
\bibinfo{author}{\bibfnamefont{L.-J.} \bibnamefont{Wang}},
  \bibinfo{author}{\bibfnamefont{J.}~\bibnamefont{Engel}}, \bibnamefont{and}
  \bibinfo{author}{\bibfnamefont{J.~M.} \bibnamefont{Yao}},
  \bibinfo{journal}{Phys. Rev. C} \textbf{\bibinfo{volume}{98}},
  \bibinfo{pages}{031301} (\bibinfo{year}{2018}).

\bibitem[{\citenamefont{Coraggio et~al.}(2019)\citenamefont{Coraggio, Gargano,
  Itaco, and Nowacki}}]{Coraggio19b}
\bibinfo{author}{\bibfnamefont{L.}~\bibnamefont{Coraggio}},
  \bibinfo{author}{\bibfnamefont{A.}~\bibnamefont{Gargano}},
  \bibinfo{author}{\bibfnamefont{N.}~\bibnamefont{Itaco}}, \bibnamefont{and}
  \bibinfo{author}{\bibfnamefont{F.}~\bibnamefont{Nowacki}}
  (\bibinfo{year}{2019}), \bibinfo{note}{in preparation}.

\end{thebibliography}

\newpage

\appendix*
\renewcommand{\thetable}{A.\Roman{table}}
\setcounter{table}{0}
\section{Tables of SP energies and effective operators}
\subsection{SP energies}

\begin{table}[H]
\caption{Theoretical proton and neutron SP energy
  spacings (in MeV) for $^{40}$Ca core.}
\begin{ruledtabular}
\begin{tabular}{ccc}
\label{spe40Ca}
 ~ & Proton SP spacings & Neutron SP spacings \\
\colrule
 $0f_{7/2}$   & 0.0 & 0.0 \\ 
 $0f_{5/2}$   & 8.6 & 7.8 \\ 
 $1p_{3/2}$   & 1.6 & 2.1 \\ 
 $1p_{1/2}$   & 3.3 & 4.0 \\ 
\end{tabular}
\end{ruledtabular}
\end{table}

\begin{table}[H]
\caption{Theoretical proton and neutron SP energy
  spacings (in MeV) for $^{56}$Ni core.}
\begin{ruledtabular}
\begin{tabular}{ccc}
\label{spe56Ni}
 ~ & Proton SP spacings & Neutron SP spacings \\
\colrule
 $0f_{5/2}$   & 0.2 & 0.0 \\ 
 $1p_{3/2}$   & 0.0 & 0.5 \\ 
 $1p_{1/2}$   & 0.6 & 1.1 \\ 
 $0g_{9/2}$   & 3.1 & 3.5 \\ 
\end{tabular}
\end{ruledtabular}
\end{table}

\begin{table}[H]
\caption{Theoretical proton and neutron SP energy
  spacings (in MeV) for $^{100}$Sn core.}
\begin{ruledtabular}
\begin{tabular}{ccc}
\label{spe100Sn}
 ~ & Proton SP spacings & Neutron SP spacings \\
\colrule
 $0g_{7/2}$   & 0.0 & 0.0 \\ 
 $1d_{5/2}$   & 0.3 & 0.6 \\ 
 $1d_{3/2}$   & 1.2 & 1.5 \\ 
 $2s_{1/2}$   & 1.1 & 1.2 \\ 
 $0h_{11/2}$ & 1.9 & 2.7 \\ 
\end{tabular}
\end{ruledtabular}
\end{table}

\subsection{Effective $M1$ and GT operators}

\begin{table}[H]
\caption{Proton and neutron matrix elements of the effective magnetic
  dipole operator $M1$ (I) (in $\mu_N$) for the $^{40}$Ca core. In the
  last column we report the corresponding quenching
  factors. For $l-$forbidden matrix elements there is no quenching
  factor to be shown.} 
\begin{ruledtabular}
\begin{tabular}{cccc}
\label{effM1_40Ca}
$n_a l_a j_a ~ n_b l_b j_b $ & $T_z$ & $M1_{\rm eff}$ &  quenching factor \\
\colrule
 $0f_{7/2}~0f_{7/2}$     & +1/2  & 8.760 & 0.965 \\ 
 $0f_{7/2}~0f_{5/2}$     & +1/2  & -3.986 & 0.961 \\ 
 $0f_{5/2}~0f_{7/2}$     & +1/2  & 4.640 & 1.118 \\ 
 $0f_{5/2}~0f_{5/2}$     & +1/2  & 1.310 & 1.073 \\ 
 $0f_{5/2}~1p_{3/2}$     & +1/2  & -0.017 &   ~ \\ 
 $1p_{3/2}~0f_{5/2}$     & +1/2  &  0.014 &   ~ \\ 
 $1p_{3/2}~1p_{3/2}$     & +1/2  & 4.462 & 0.933 \\
 $1p_{3/2}~1p_{1/2}$     & +1/2  & -2.396 & 0.926 \\
 $1p_{1/2}~1p_{3/2}$     & +1/2  &  2.377 & 0.919 \\ 
 $1p_{1/2}~1p_{1/2}$     & +1/2  & -0.304 & 0.962 \\
 $0f_{7/2}~0f_{7/2}$     & -1/2  & -2.237 & 0.746 \\ 
 $0f_{7/2}~0f_{5/2}$     & -1/2  &  3.308 & 0.956 \\ 
 $0f_{5/2}~0f_{7/2}$     & -1/2  & -3.582 & 1.035 \\ 
 $0f_{5/2}~0f_{5/2}$     & -1/2  & 2.727 & 1.409 \\ 
 $0f_{5/2}~1p_{3/2}$     & -1/2  & -0.026 &   ~ \\ 
 $1p_{3/2}~0f_{5/2}$     & -1/2  &  0.024 &   ~ \\ 
 $1p_{3/2}~1p_{3/2}$     & -1/2  & -2.074 & 0.859 \\
 $1p_{3/2}~1p_{1/2}$     & -1/2  &  2.025 & 0.938 \\
 $1p_{1/2}~1p_{3/2}$     & -1/2  & -2.008 & 0.930 \\ 
 $1p_{1/2}~1p_{1/2}$     & -1/2  &  0.799 & 1.047 \\
\end{tabular}
\end{ruledtabular}
\end{table}

\begin{table}[H]
\caption{Same as in Table \ref{effM1_40Ca}, but for the $^{56}$Ni core.}
\begin{ruledtabular}
\begin{tabular}{cccc}
\label{effM1_56Ni}
$n_a l_a j_a ~ n_b l_b j_b $ & $T_z$ & $M1_{\rm eff}$ &  quenching factor \\
\colrule
 $0f_{5/2}~0f_{5/2}$     & +1/2  & 2.212 & 1.812 \\ 
 $0f_{5/2}~1p_{3/2}$     & +1/2  & -0.033 &   ~ \\ 
 $1p_{3/2}~0f_{5/2}$     & +1/2  &  0.026 &   ~ \\ 
 $1p_{3/2}~1p_{3/2}$     & +1/2  & 3.358 & 0.739 \\
 $1p_{3/2}~1p_{1/2}$     & +1/2  & -1.554 & 0.601 \\
 $1p_{1/2}~1p_{3/2}$     & +1/2  &  1.586 & 0.613 \\ 
 $1p_{1/2}~1p_{1/2}$     & +1/2  & -0.091 & 0.288 \\
 $0g_{9/2}~0g_{9/2}$     & +1/2  & 10.174 & 0.877 \\ 
 $0f_{5/2}~0f_{5/2}$     & -1/2  & 1.338 & 0.691 \\ 
 $0f_{5/2}~1p_{3/2}$     & -1/2  & -0.024 &   ~ \\ 
 $1p_{3/2}~0f_{5/2}$     & -1/2  &  0.028 &   ~ \\ 
 $1p_{3/2}~1p_{3/2}$     & -1/2  & -1.233 & 0.511 \\
 $1p_{3/2}~1p_{1/2}$     & -1/2  &  1.178 & 0.546 \\
 $1p_{1/2}~1p_{3/2}$     & -1/2  & -1.209 & 0.560 \\ 
 $1p_{1/2}~1p_{1/2}$     & -1/2  &  0.512 & 0.671 \\
 $0g_{9/2}~0g_{9/2}$     & -1/2  & -0.473 & 0.145 \\ 
\end{tabular}
\end{ruledtabular}
\end{table}

\begin{table}[H]
\caption{Same as in Table \ref{effM1_40Ca}, but for the $^{100}$Sn core.} 
\begin{ruledtabular}
\begin{tabular}{cccc}
\label{effM1_100Sn}
$n_a l_a j_a ~ n_b l_b j_b $ & $T_z$ & $M1_{\rm eff}$ &  quenching factor \\
\colrule
$0g_{7/2}~0g_{7/2}$     & +1/2  & 3.013 & 1.120 \\ 
$0g_{7/2}~1d_{5/2}$     & +1/2  & -0.064 &   ~ \\ 
$1d_{5/2}~0g_{7/2}$     & +1/2  & 0.060 &   ~ \\ 
$1d_{5/2}~1d_{5/2}$     & +1/2  & 5.190 & 0.765\\ 
$1d_{5/2}~1d_{3/2}$     & +1/2  & -2.180 & 0.628 \\ 
$1d_{3/2}~1d_{5/2}$     & +1/2  &  2.274 & 0.655 \\ 
$1d_{3/2}~1d_{3/2}$     & +1/2  & 0.407 & 2.599 \\ 
$1d_{3/2}~2s_{1/2}$     & +1/2  & -0.123 &   ~ \\ 
$2s_{1/2}~1d_{3/2}$     & +1/2  & 0.119 &   ~ \\ 
$2s_{1/2}~2s_{1/2}$     & +1/2  & 2.453 & 0.734 \\ 
$0h_{11/2}~0h_{11/2}$ & +1/2  & 12.349 & 0.861 \\ 
$0g_{7/2}~0g_{7/2}$     & -1/2  & 1.984 & 0.851 \\ 
$0g_{7/2}~1d_{5/2}$     & -1/2  & -0.008 &   ~ \\ 
$1d_{5/2}~0g_{7/2}$     & -1/2  & 0.008 &   ~ \\ 
$1d_{5/2}~1d_{5/2}$     & -1/2  & -1.417 & 0.523 \\ 
$1d_{5/2}~1d_{3/2}$     & -1/2  & 1.681 & 0.580 \\ 
$1d_{3/2}~1d_{5/2}$     & -1/2  & -1.756 & 0.606 \\ 
$1d_{3/2}~1d_{3/2}$     & -1/2  & 1.081 & 0.746 \\ 
$1d_{3/2}~2s_{1/2}$     & -1/2  & 0.076 &   ~ \\ 
$2s_{1/2}~1d_{3/2}$     & -1/2  & -0.071 &   ~ \\ 
$2s_{1/2}~2s_{1/2}$     & -1/2  & -1.414 & 0.618 \\ 
$0h_{11/2}~0h_{11/2}$ & -1/2  & -0.696 & 0.198 \\ 
\end{tabular}
\end{ruledtabular}
\end{table}

\begin{table}[H]
\caption{Matrix elements of the proton-neutron effective GT$^+$ and
  GT$^-$ operators for the $^{40}$Ca core. In the last two columns we
  report the corresponding quenching factors of present work (I) and
  those reported in the work of Ref. \cite{Siiskonen01} (see text for
  details). For $l-$forbidden matrix elements there is no quenching
  factor to be shown.}
\begin{ruledtabular}
\begin{tabular}{cccc}
\label{effGT_40Ca}
$n_a l_a j_a ~ n_b l_b j_b $ & GT$^-_{\rm eff}$ &  quenching
factor \\
\colrule
 $0f_{7/2}~0f_{7/2}$     &  2.870 & 0.995 \\ 
 $0f_{7/2}~0f_{5/2}$     & -3.210 & 0.964 \\ 
 $0f_{5/2}~0f_{7/2}$     &  3.941 & 1.183 \\ 
 $0f_{5/2}~0f_{5/2}$     & -2.104 & 1.130 \\ 
 $0f_{5/2}~1p_{3/2}$    & -0.033 &   ~ \\ 
 $1p_{3/2}~0f_{5/2}$    &   0.001 &   ~ \\ 
 $1p_{3/2}~1p_{3/2}$   &  2.162 & 0.931 \\
 $1p_{3/2}~1p_{1/2}$   & -1.906 & 0.918 \\
 $1p_{1/2}~1p_{3/2}$   &   1.901 & 0.915 \\ 
 $1p_{1/2}~1p_{1/2}$   & -0.691 & 0.953 \\
\colrule
$n_a l_a j_a ~ n_b l_b j_b $ & GT$^+_{\rm eff}$ &  quenching
factor (I) &  quenching factor (II)\\
\colrule
 $0f_{7/2}~0f_{7/2}$     &  2.706 & 0.938 & 0.905 \\ 
 $0f_{7/2}~0f_{5/2}$     & -3.012 & 0.904 & 0.856 \\ 
 $0f_{5/2}~0f_{7/2}$     &  3.276 & 0.984 &  ~ \\ 
 $0f_{5/2}~0f_{5/2}$     & -1.737 & 0.932 & 0.882 \\ 
 $0f_{5/2}~1p_{3/2}$    & -0.001 &   ~ &   ~ \\ 
 $1p_{3/2}~0f_{5/2}$    &   0.026 &   ~ &   ~ \\ 
 $1p_{3/2}~1p_{3/2}$   &   2.135 & 0.921 & 0.880 \\
 $1p_{3/2}~1p_{1/2}$   & -1.879 & 0.904 & 0.863 \\
 $1p_{1/2}~1p_{3/2}$   &   1.871 & 0.901 &  ~ \\ 
 $1p_{1/2}~1p_{1/2}$   & -0.686 & 0.935 & 0.932 \\
\end{tabular}
\end{ruledtabular}
\end{table}

\begin{table}[H]
\caption{ Same as in Table \ref{effGT_40Ca}, but for the $^{56}$Ni
  core.} 
\begin{ruledtabular}
\begin{tabular}{cccc}
\label{effGT_56Ni}
$n_a l_a j_a ~ n_b l_b j_b $ & GT$^-_{\rm eff}$ &  quenching
factor \\
\colrule
 $0f_{5/2}~0f_{5/2}$     & -0.674 & 0.362 \\ 
 $0f_{5/2}~1p_{3/2}$    & -0.085 &   ~ \\ 
 $1p_{3/2}~0f_{5/2}$    &   0.006 &   ~ \\ 
 $1p_{3/2}~1p_{3/2}$   &  1.441 & 0.620 \\
 $1p_{3/2}~1p_{1/2}$   & -1.141 & 0.549 \\
 $1p_{1/2}~1p_{3/2}$   &   1.189 & 0.572 \\ 
 $1p_{1/2}~1p_{1/2}$   & -0.482 & 0.657 \\
 $0g_{9/2}~0g_{9/2}$     &  1.608 & 0.511\\ 
\colrule
$n_a l_a j_a ~ n_b l_b j_b $ & GT$^+_{\rm eff}$ &  quenching
factor (I) &  quenching factor (II)\\
\colrule
 $0f_{5/2}~0f_{5/2}$     & -0.638 & 0.342 & 0.458 \\ 
 $0f_{5/2}~1p_{3/2}$    & -0.011 &   ~ &   ~ \\ 
 $1p_{3/2}~0f_{5/2}$    &   0.061 &   ~ &   ~ \\ 
 $1p_{3/2}~1p_{3/2}$   &  1.405 & 0.605 & 0.689 \\
 $1p_{3/2}~1p_{1/2}$   & -1.159 & 0.558 & 0.680 \\
 $1p_{1/2}~1p_{3/2}$   &   1.121 & 0.539 & ~  \\ 
 $1p_{1/2}~1p_{1/2}$   & -0.468 & 0.638 & ~  \\
 $0g_{9/2}~0g_{9/2}$     &  1.536 & 0.488 & 0.802 \\ 
\end{tabular}
\end{ruledtabular}
\end{table}

\begin{table}[H]
\caption{ Same as in Table \ref{effGT_40Ca}, but for the $^{100}$Sn
  core.} 
\begin{ruledtabular}
\begin{tabular}{cccc}
\label{effGT_100Sn}
$n_a l_a j_a ~ n_b l_b j_b $ & GT$^-_{\rm eff}$ &  quenching
factor \\
\colrule
 $0g_{7/2}~0g_{7/2}$     & -1.168 & 0.521 \\ 
 $0g_{7/2}~1d_{5/2}$     & -0.108 & ~ \\ 
 $1d_{5/2}~0g_{7/2}$     & 0.000 & ~ \\ 
 $1d_{5/2}~1d_{5/2}$     & 1.686 & 0.647 \\ 
 $1d_{5/2}~1d_{3/2}$     & -1.525 & 0.547 \\ 
 $1d_{3/2}~1d_{5/2}$     & 1.708 & 0.613 \\ 
 $1d_{3/2}~1d_{3/2}$     & -0.888 & 0.638 \\ 
 $1d_{3/2}~2s_{1/2}$     & -0.124 & ~ \\ 
 $2s_{1/2}~1d_{3/2}$     & 0.093 & ~ \\ 
 $2s_{1/2}~2s_{1/2}$     & 1.405 & 0.638 \\ 
 $0h_{11/2}~0h_{11/2}$  & 1.931 & 0.570 \\ 
\colrule
$n_a l_a j_a ~ n_b l_b j_b $ & GT$^+_{\rm eff}$ &  quenching
factor (I) &  quenching factor (II) \\
\colrule
 $0g_{7/2}~0g_{7/2}$     & -1.168 & 0.521 & 0.472 \\ 
 $0g_{7/2}~1d_{5/2}$     & 0.001 & ~ & ~ \\ 
 $1d_{5/2}~0g_{7/2}$     & 0.102 & ~ & ~ \\ 
 $1d_{5/2}~1d_{5/2}$     & 1.686 & 0.647 & 0.595 \\ 
 $1d_{5/2}~1d_{3/2}$     & -1.688 & 0.606 & 0.513 \\ 
 $1d_{3/2}~1d_{5/2}$     & 1.543 & 0.553 & ~ \\ 
 $1d_{3/2}~1d_{3/2}$     & -0.888 & 0.638 & 0.652 \\ 
 $1d_{3/2}~2s_{1/2}$     & -0.098 & ~ & ~ \\ 
 $2s_{1/2}~1d_{3/2}$     & 0.117 & ~ & ~ \\ 
 $2s_{1/2}~2s_{1/2}$     & 1.405 & 0.638 & ~ \\ 
 $0h_{11/2}~0h_{11/2}$  & 1.931 & 0.570 & ~ \\ 
\end{tabular}
\end{ruledtabular}
\end{table}

\end{document}